\documentclass[american]{article}
\usepackage[T1]{fontenc}
\usepackage[utf8]{inputenc}
\usepackage{xcolor}
\usepackage{babel}
\usepackage{float}
\usepackage{amsmath}
\usepackage{amssymb}
\usepackage{graphicx}
\usepackage[a4paper]{geometry}
\usepackage[pdfusetitle,
 bookmarks=true,bookmarksnumbered=false,bookmarksopen=false,
 breaklinks=false,pdfborder={0 0 1},backref=false,colorlinks=true]
 {hyperref}
\hypersetup{
 linkcolor=blue,citecolor=blue,urlcolor=blue,pdfauthor={Simona Beresova, Michal Beres, Tomas Luber, Stanislav Sysala}}

\makeatletter

\providecommand{\tabularnewline}{\\}
\floatstyle{ruled}
\newfloat{algorithm}{tbp}{loa}
\providecommand{\algorithmname}{Algorithm}
\floatname{algorithm}{\protect\algorithmname}

\numberwithin{equation}{section}
\numberwithin{figure}{section}
\numberwithin{table}{section}

\usepackage{xcolor}
\IfFileExists{algpseudocode.sty}{
\usepackage{algpseudocode}
\algrenewcommand\algorithmicrequire{\textbf{Given:}}
}{
\newlength{\algorithmicindent}
\newlength{\algorithmiccurrentindent}
\setlength{\algorithmicindent}{1.5em}
\newcounter{algorithmicline}
\newenvironment{algorithmic}[1][0]{
	\setcounter{algorithmicline}{0}
	\setlength{\algorithmiccurrentindent}{0pt}
	\begin{list}{}{
		\setlength{\leftmargin}{0pt}
		\setlength{\labelwidth}{2em}
		\setlength{\labelsep}{0.5em}
		\setlength{\itemindent}{0pt}
		\setlength{\itemsep}{0.2ex}
		\setlength{\parsep}{0pt}
		\setlength{\topsep}{0.2ex}
		\setlength{\partopsep}{0pt}
	}
}{
	\end{list}
}
\newcommand{\algorithmicrequire}{\textbf{Given:}}
\newcommand{\State}{
	\stepcounter{algorithmicline}
	\item[]{\arabic{algorithmicline}\hspace*{\algorithmiccurrentindent}\makebox[2em][l]~ }
}
\newcommand{\Statex}{\item[]\hspace*{\algorithmiccurrentindent}}
\newcommand{\Require}{\State \algorithmicrequire\ }
\newcommand{\For}[1]{\State \textbf{for} ##1 \textbf{~}\addtolength{\algorithmiccurrentindent}{\algorithmicindent}}
\newcommand{\EndFor}{\addtolength{\algorithmiccurrentindent}{-\algorithmicindent}\State \textbf{end for}}
\newcommand{\If}[1]{\State \textbf{if} ##1 \textbf{~}\addtolength{\algorithmiccurrentindent}{\algorithmicindent}}
\newcommand{\Else}{\addtolength{\algorithmiccurrentindent}{-\algorithmicindent}\State \textbf{else}\addtolength{\algorithmiccurrentindent}{\algorithmicindent}}
\newcommand{\EndIf}{\addtolength{\algorithmiccurrentindent}{-\algorithmicindent}\State \textbf{end if}}
}
\definecolor{revisionpurple}{RGB}{0,0,0}
\ifdefined\showcaptionsetup
 \PassOptionsToPackage{caption=false}{subfig}
\fi
\usepackage{subfig}
\makeatother

\begin{document}
\title{Delayed acceptance sampling with Hamiltonian proposal subchains for
random field materials inference}
\author{Simona Bérešová$^{1,2}$, Michal Béreš$^{1,2}$, Tomáš Luber$^{1}$,
and Stanislav Sysala$^{1}$\\[0.4em]\footnotesize$^{1}$ Institute
of Geonics of the Czech Academy of Sciences, Ostrava, Czech Republic\\[-0.1em]\footnotesize$^{2}$
Faculty of Electrical Engineering and Computer Science,\\[-0.1em]\footnotesize
VŠB - Technical University of Ostrava, Ostrava, Czech Republic}
\maketitle
\begin{abstract}
This paper focuses on accelerating Markov chain Monte Carlo (MCMC)
sampling in Bayesian inverse problems in which forward model evaluations
dominate the computational cost. It builds on several established
ingredients that have previously been used successfully in related
scenarios: delayed acceptance, neural network surrogate models, Hamiltonian
proposals, and proposal subchains. The main algorithmic framework
is the delayed-acceptance Metropolis-Hastings algorithm of Christen
and Fox (2005). The first-stage proposal distribution is constructed
from a subchain of Hamiltonian trajectories targeting the surrogate
posterior. For each fixed surrogate model, the Hamiltonian subchain
and delayed-acceptance correction define a kernel invariant with respect
to the exact posterior. In the present work, the surrogate is updated
only during a burn-in phase, after which the production run uses a
fixed surrogate model.

The sampling framework is implemented in Python using parallel processes.
Several chains are generated in parallel and share a single surrogate
model trained during the burn-in phase on all collected data. The
forward model is treated as a black box; therefore, the application
area is broad. However, the main motivation behind this research is
the efficient solution of geotechnical inverse problems with material
properties represented by Gaussian random fields. In this study, the
sampling framework is applied to a geotechnical inverse problem in
which hydraulic conductivity and porosity are modeled as non-stationary
Gaussian random fields (GRFs) approximated using a truncated Karhunen-Loeve
expansion. Based on a precomputation, the truncation dimensions of
the Karhunen-Loeve expansions are chosen separately for hydraulic
conductivity and porosity. The forward model outputs are pore pressure
values at the control points and selected observation times. These
are compared with in situ pore pressure measurements collected over
one year during the Tunnel Sealing Experiment conducted in an underground
laboratory in Canada.
\end{abstract}

\section{Introduction}

Bayesian inversion has become an important tool in many areas of applied
mathematics. It provides a natural way to integrate observational
data with prior knowledge in order to quantify uncertainties in the
unknown parameters via the posterior distribution. For posterior sampling,
Markov chain Monte Carlo (MCMC) sampling methods are typical tools.
In particular, the Metropolis--Hastings (MH) algorithm provides a
robust mechanism for generating samples from complex posterior distributions.
The MH algorithm is conceptually simple, requires minimal parameter
tuning, and forms a simple building block for more advanced MCMC schemes.
However, a major obstacle to the practical application of the classical
MH algorithm is the computational cost of evaluating the forward model.
In many applied fields, including geotechnics, hydrology, and structural
mechanics, the forward model involves the numerical solution of a
system of partial differential equations (PDEs) or other computationally
demanding models. When each forward model evaluation requires several
seconds or minutes, naive MCMC becomes infeasible, especially in combination
with high-dimensional unknown parameters.

Bayesian inversion with computationally expensive forward models has
motivated a significant effort on the development of Markov chain
Monte Carlo methods that reduce the number of exact model evaluations
without sacrificing posterior correctness. A central idea in this
literature is the delayed acceptance Metropolis-Hastings algorithm,
introduced by Christen and Fox \cite{christen_markov_2005}, where
a cheap approximation is used to screen proposals before the expensive
exact model is evaluated. The advantage of this approach is that it
preserves the exact target distribution through a second-stage correction
while reducing the cost of unsuccessful proposals.

Much effort has also been given to the use of various surrogate models
for MCMC acceleration. For example, \cite{efendiev_preconditioning_2006}
showed that coarse models can be used to precondition sampling in
expensive subsurface problems. Later, \cite{cui_bayesian_2011} proposed
an adaptive delayed-acceptance strategy, demonstrating that an approximate
forward model can be updated during sampling while the algorithm targets
the exact posterior, and used this framework for geothermal reservoir
calibration.\textcolor{revisionpurple}{~For general-purpose settings with no obvious coarse model, \cite{sherlock_adaptive_2017} propose MCMC with an adaptive construction of a posterior approximation based on a weighted average of previous evaluations.}

Another relevant topic concerns gradient-informed proposals, especially
Hamiltonian Monte Carlo, see e.g.~\cite{neal_mcmc_2011}. Hamiltonian
methods were introduced in molecular simulation and later developed
into a powerful MCMC methodology because they can generate distant
proposals with relatively high acceptance probability, which substantially
reduces random-walk behavior in correlated posterior distributions.
The direction closest to the present paper is the intersection of
delayed acceptance, surrogate modeling, and Hamiltonian proposals.
The paper \cite{deveney_deep_2023} combines neural network surrogate
models, delayed acceptance, and Hamiltonian Monte Carlo for a PDE-based
Bayesian inverse problem. The present paper uses a similar delayed-acceptance
setting with one specific modification: the first-stage proposal is
a finite subchain of Hamiltonian trajectories targeting an approximate
posterior.  The subchain strategy was successfully used by \cite{lykkegaard_multilevel_2023}
within a multilevel delayed acceptance sampling scheme. Although that
work is not specifically Hamiltonian, it directly supports the construction
used here: if the inner kernel is reversible with respect to the approximate
target, then the finite inner subchain can serve as a valid first-stage
proposal in delayed acceptance.

The remaining theoretical issue is adaptation. Once the surrogate
changes during the sampling process, we are using an adaptive MCMC
algorithm, so stepwise exactness is not enough to guarantee ergodicity.\textcolor{revisionpurple}{~This area is well studied; \cite{bai_containment_2011,roberts_coupling_2007,andrieu_ergodicity_2006} provide the standard framework based on diminishing adaptation and containment conditions.}\textcolor{revisionpurple}{~In the present paper, every delayed-acceptance kernel with a frozen surrogate model targets the exact posterior. For online-trained neural network surrogates, the standard diminishing-adaptation and containment conditions are not satisfied. Therefore, the neural network is updated only during a burn-in phase, after which the surrogate model remains fixed.}
Future inspiration regarding controlled adaptation can be found e.g.
in~\cite{conrad_accelerating_2016}, which showed that changing local
approximations can still lead to asymptotically exact sampling.

In the present paper, we combine delayed acceptance, finite Hamiltonian
proposal subchains targeting a surrogate posterior distribution,  and
a neural-network surrogate model shared by several sampling processes.
The shared surrogate model can be improved using collected data during
the burn-in phase. This lets us study the method in a realistic PDE-constrained
setting. The sampling framework is used within a Bayesian solution
of a benchmark geotechnical inverse problem in which spatially varying
material properties are represented by truncated Gaussian random fields.

\section{Inverse problem in its general form}

Consider a probability space $\left(\Xi,\mathcal{F}_{\Xi},\mathbb{P}\right)$
and the measurable spaces $(\mathbb{R}^{n},\mathcal{B}(\mathbb{R}^{n}))$
and $(\mathbb{R}^{m},\mathcal{B}(\mathbb{R}^{m}))$, where $n,m\in\mathbb{N}$.
Let $G:\mathbb{R}^{n}\rightarrow\mathbb{R}^{m}$ be a measurable,
deterministic forward model mapping input parameters to model outputs.
The aim of the inverse problem is to characterize the unknown model
inputs corresponding to a given vector of noisy outputs $y\in\mathbb{R}^{m}$.

In practical applications, model outputs are not observed exactly
due to measurement noise. Therefore, the observed data are modeled
as a random variable $Y:\Omega\rightarrow\mathbb{R}^{m}$, and observational
noise is represented by a random variable $Z:\Omega\rightarrow\mathbb{R}^{m}$.
The unknown input parameters are modeled as a random variable $U:\Omega\rightarrow\mathbb{R}^{n}$.
Throughout this paper, we assume that $Z$ is independent of $U$.
The relationship between these quantities is described by the additive
noise model
\begin{equation}
Y=G(U)+Z.
\end{equation}
A realization $y\in\mathbb{R}^{m}$ of $Y$ is referred to as the
observed data.

The Bayesian inverse problem consists of determining the conditional
distribution of $U$ given the observation $Y=y$, called the posterior
distribution. In addition to the observational model, the Bayesian
approach incorporates prior information about the unknown parameters
in the form of a prior distribution, which reflects preliminary knowledge
based on experience or modeling assumptions. Bayesian inference can
thus be understood as a refinement of prior information using observed
data.

We assume that the prior distribution of $U$ and the noise distribution
of $Z$ admit probability density functions with respect to the Lebesgue
measure on $\mathbb{R}^{n}$ and $\mathbb{R}^{m}$, respectively.
These densities are denoted by $f_{U}$ and $f_{Z}$, and are referred
to as the prior pdf and the noise pdf. The conditional pdf of $Y$
given $U=u$ is denoted by $f_{Y|U}(y|u)$. For fixed $y$, the function
$u\mapsto f_{Y|U}(y|u)$ is called the data likelihood.

Under the additive noise model assumption, the conditional density
satisfies
\begin{equation}
f_{Y|U}(y|u)=f_{Z}\bigl(y-G(u)\bigr).
\end{equation}
Given the observed data $y$, Bayes’ theorem yields the posterior
density
\begin{equation}
f_{U|Y}(u|y)=\frac{f_{Y|U}(y|u)f_{U}(u)}{\int_{\mathbb{R}^{n}}f_{Y|U}(y|v)f_{U}(v)\,\mathrm{d}v}=\frac{f_{Z}\bigl(y-G(u)\bigr)f_{U}(u)}{\int_{\mathbb{R}^{n}}f_{Z}\bigl(y-G(v)\bigr)f_{U}(v)\,\mathrm{d}v}.
\end{equation}
We assume that the normalizing constant in the denominator is finite,
ensuring that the posterior distribution is well defined. Using proportionality
notation, the posterior density can be written as
\begin{equation}
f_{U|Y}(u|y)\propto\underbrace{f_{Z}\bigl(y-G(u)\bigr)}_{\text{likelihood}}\underbrace{f_{U}(u)}_{\text{prior}}\label{eq:posterior-1}
\end{equation}
almost everywhere with respect to the Lebesgue measure on $\mathbb{R}^{n}$.

Obtaining an explicit expression for the posterior density is only
the first step in solving a Bayesian inverse problem. The main challenge
is to make this expression computationally useful, for example by
generating samples from the posterior distribution. Basic Monte Carlo
sampling is not possible since the normalizing constant is unknown.
Therefore, samples are typically generated using Markov chain Monte
Carlo (MCMC) methods, which are described in~\textcolor{revisionpurple}{Section}~\ref{sec:Sampling-framework}.

\section{Geotechnical benchmark problem}\label{sec:Geotechnical-benchmark-problem}

The geotechnical application considered in this paper is based on
the tunnel-sealing experiment (TSX), performed in an underground research
laboratory in Canada in 1997 and 1998; for a detailed description
of this experiment, see the technical report \cite{chandler_five_2002}.
It has previously been used as a benchmark for both forward and inverse
modeling, see, e.g., \cite{rutqvist_modeling_2009,Luber2024,Beresova2022Bayesian}.The
experiment was designed to allow two-dimensional modeling.

\begin{figure}[H]
\centering{}\subfloat[\label{fig:TSX_problem_geometry}Problem geometry]{\includegraphics[width=0.46\textwidth]{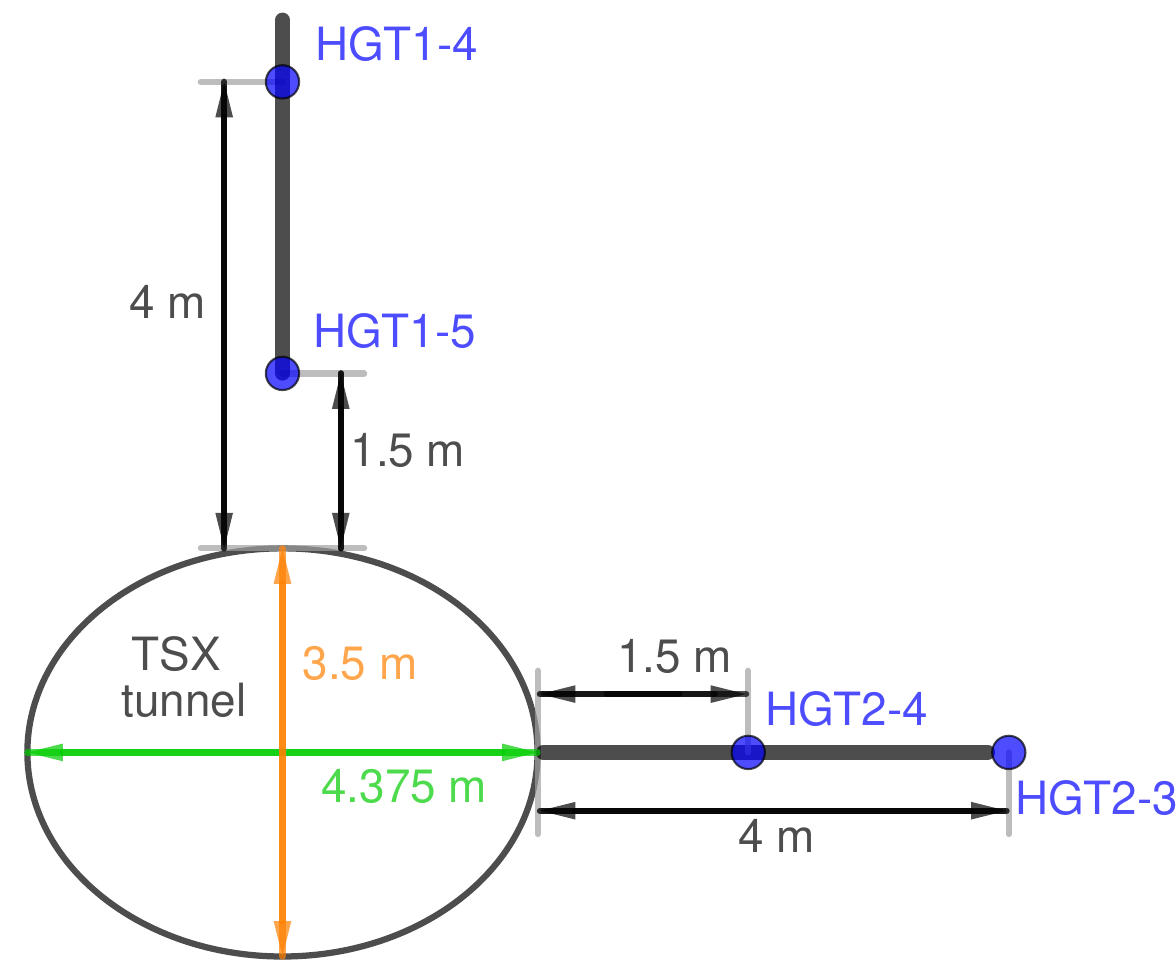}

}\subfloat[\label{fig:TSX_measurements}Pore pressure measurements]{\includegraphics[width=0.46\textwidth]{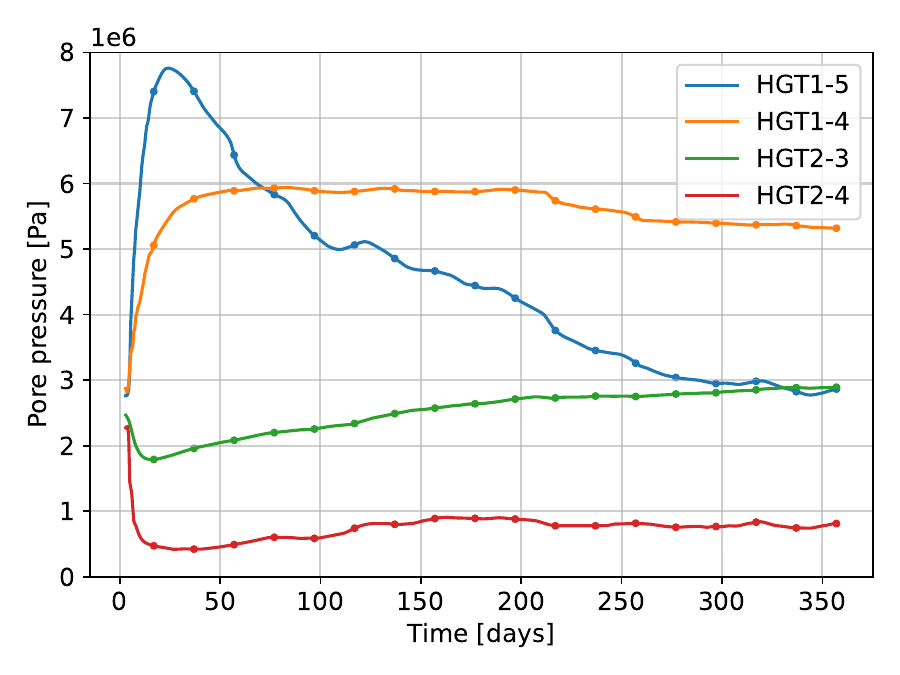}

}\caption{Problem geometry (tunnel cross-section and location of control points)
and in-situ measurements (time series of pore pressure measurements
at the control points)}
\end{figure}

Figure~\ref{fig:TSX_problem_geometry} shows the problem geometry:
the tunnel cross-section and the locations of the control points.
The available data consist of four time series representing pore pressure
measurements at these control points over one year, see Fig.~\ref{fig:TSX_measurements}.
\textcolor{revisionpurple}{For the purposes of the inverse problem, each time series was sampled at 18 uniformly distributed time points represented by dots in Fig.~\ref{fig:TSX_measurements}. These samples define the observed data \(y\in\mathbb{R}^{72}\). Observational noise was chosen to be additive Gaussian with independent components \(\mathcal{N}(0,(20\rho g)^2)\). The constants \(\rho\) and \(g\) provide the conversion from water-column height in meters to pressure in Pascals; they are defined in Table~\ref{tab:Fixed-material-properties}.}

\subsection{PDE model}

For the modeling of the hydro-mechanical processes around the tunnel,
we consider the two-dimensional Biot poroelasticity model with zero
volume forces given by equations
\begin{align}
-\text{div}\left(C:\varepsilon\left(u\right)\right)+{\color{revisionpurple}\alpha_{B}\nabla p} & =0\label{eq: mechanical_balance}\\
S\frac{\partial p}{\partial t}-\nabla\cdot\left(\frac{K}{\rho g}\nabla p\right)+\alpha_{B}\frac{\partial}{\partial t}\left(\text{div}\left(u\right)\right) & =0\label{eq: mass_balance}
\end{align}
in $\Omega\subset\mathbb{R}^{2}$. Here, $u$ is the displacement
vector (in meters), $p$ is pore pressure in Pascals, $C$ is the
elasticity tensor, $\varepsilon$ is the small strain tensor
\[
\varepsilon\left(u\right)=\frac{1}{2}\left(\nabla u+\left(\nabla u\right)^{T}\right),
\]
and the elastic effective-stress term is
\[
C:\varepsilon\left(u\right)=2\mu\varepsilon\left(u\right)+\lambda\text{div}\left(u\right)I
\]
 for an isotropic linear elastic medium; $I$ is the identity tensor.{}
Material parameters $\alpha_{B}$, $S$, $K$, $\mu$, $\lambda$
and constants $\rho$, $g$ will be specified in~\textcolor{revisionpurple}{Section}~\ref{subsec:Parameterization-of-inputs}.

\begin{figure}[H]
\begin{centering}
\subfloat[whole domain]{\includegraphics[width=0.46\textwidth]{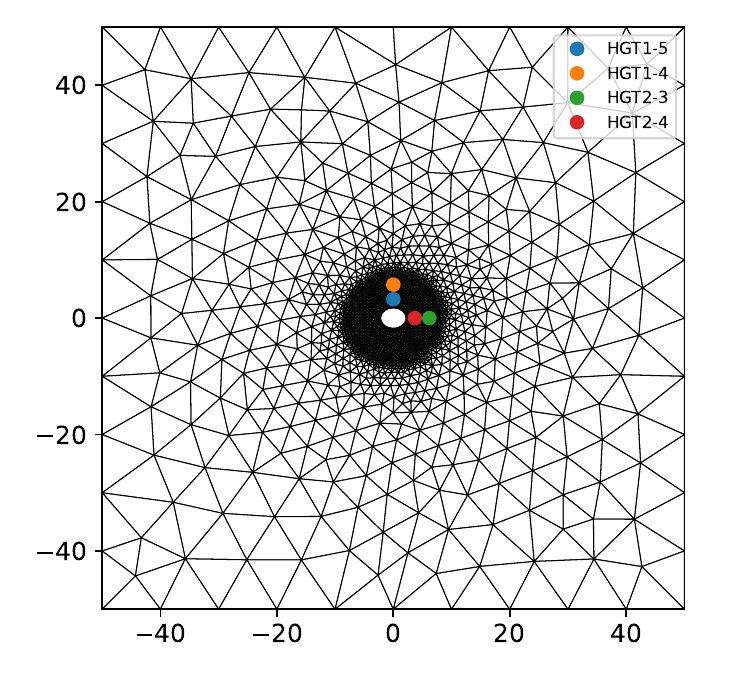}

}\subfloat[cutout $(-10,10)\times(-10,10)$]{\includegraphics[width=0.46\textwidth]{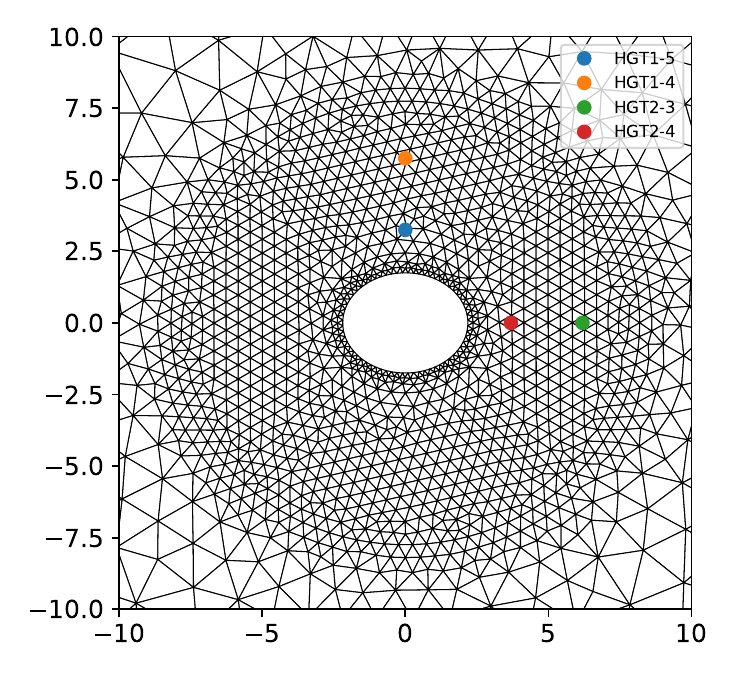}

}
\par\end{centering}
\caption{Discretization of $\Omega=(-50,50)\times(-50,50)$}\label{fig:Mesh}

\end{figure}

The domain $\Omega$ is a square $(-50,50)\times(-50,50)$ with an
elliptic hole representing the tunnel cross-section, see Fig.~\ref{fig:Mesh}.
The ellipse is placed in the center of the square and its axes are
parallel to the sides of the square. The lengths of the semi-axes
are $r_{x}=4.375/2$~m and $r_{y}=3.5/2$~m.  An unstructured mesh
with $N=1884$ nodes and $3664$ triangular elements was chosen, and
the time step is 0.5 day during the first 17 days and 2 days during
the remaining part of the simulation (after excavation). This discretization
is a compromise between accuracy and computational cost, and allows
us to perform extensive numerical experiments in Section~\ref{sec:Efficiency-analysis}
for the sampling efficiency analysis.

We prescribe zero initial displacement $u_{0}=0$~m, initial pore
pressure $p_{0}=4.1$~MPa, and the initial stress tensor given by
the values $\sigma_{x}=45$~MPa (stress close to horizontal), $\sigma_{y}=12.8$~MPa
(stress close to vertical) and angle $\omega=8^{\circ}$ (representing
rotation from the horizontal direction). These values are based on
the technical report \cite{chandler_five_2002}. The boundary conditions
prescribed on the inner elliptic hole simulate tunnel excavation during
the first 17 days and the subsequent relaxation period (days 18 -
357). During the drilling period, the initial pore pressure and the
initial stress transformed into the normal direction decrease linearly
to zero. During the relaxation period, the pore pressure and the normal
stress are equal to zero. Pore pressure on the outer square boundary
is set to the ambient pore pressure value $p_{0}=4.1$~MPa.

From the results, we extract pore pressure values at the locations
of the four control points and at the chosen time points corresponding
to the vector of reference observations $y\in\mathbb{R}^{72}$.

\subsection{Input material properties and their prior distribution}\label{subsec:Parameterization-of-inputs}

We consider uncertainty in the following material properties: hydraulic
conductivity $K$, porosity $\phi$, drained bulk modulus $K_{d}$,
shear modulus $\mu$, and grain bulk modulus $K_{s}$; see Table~\ref{tab:Unknown-material-properties}.
The prior distributions are chosen so that the parameter means and
variances are consistent with the technical report \cite{chandler_five_2002},
see the rest of this section. Fixed material properties and parameters
of initial and boundary conditions are summarized in Table~\ref{tab:Fixed-material-properties}.
Table~\ref{tab:Derived-material-properties} summarizes the remaining
derived material properties that directly enter the PDEs.

\begin{table}[H]
\begin{centering}
\begin{tabular}{|c|c|c|c|c|}
\hline
\multicolumn{2}{|c|}{name and notation} & unit & model & parameters\tabularnewline
\hline
\hline
hydraulic conductivity & $K\left(x\right)$ & $\text{m}\cdot\text{s}^{-1}$ & (\ref{eq:field_Kh}) & $\xi^{K}_{1},\ldots\xi^{K}_{M_{K}}$\tabularnewline
\hline
porosity & $\phi\left(x\right)$ & - & (\ref{eq:field_por}) & $\xi^{\phi}_{1},\ldots\xi^{\phi}_{M_{\phi}}$\tabularnewline
\hline
drained bulk modulus & $K_{d}\left(x\right)$ & Pa & (\ref{prior_Kd}) & $K^{\text{inner}}_{d}$, $K^{\text{outer}}_{d}$\tabularnewline
\hline
shear modulus & $\mu$$\left(x\right)$ & Pa & (\ref{prior_shear}) & $\mu^{\text{inner}}$, $\mu^{\text{outer}}$\tabularnewline
\hline
grain bulk modulus & $K_{s}$ & Pa & \multicolumn{2}{c|}{spatially constant}\tabularnewline
\hline
\end{tabular}
\par\end{centering}
\caption{Unknown material properties and their prior distribution, defined
by $M_{K}+M_{\phi}+5$ independent random variables; $x$ is the spatial
variable}\label{tab:Unknown-material-properties}
\end{table}

\begin{table}[H]
\begin{centering}
\begin{tabular}{|c|c|}
\hline
name & symbol, value, unit\tabularnewline
\hline
\hline
fluid density & $\rho=1000$~$\text{kg}\cdot\text{m}^{-3}$\tabularnewline
\hline
gravitational acceleration & $g=9.81$~$\text{m}\cdot\text{s}^{-2}$\tabularnewline
\hline
stress, close to horizontal & $\sigma_{x}=45$~MPa\tabularnewline
\hline
stress, close to vertical & $\sigma_{y}=12.8$~MPa\tabularnewline
\hline
stress angle (rotation from horizontal) & $\omega=8^{\circ}$\tabularnewline
\hline
initial (ambient) pore pressure & $p_{0}=4.1$~MPa\tabularnewline
\hline
fluid bulk modulus & $K_{f}=2.15$~GPa\tabularnewline
\hline
\end{tabular}
\par\end{centering}
\caption{Fixed material properties and parameters of initial/boundary conditions,
based on the technical report \cite{chandler_five_2002}}\label{tab:Fixed-material-properties}
\end{table}

\begin{table}[H]
\begin{centering}
\begin{tabular}{|c|c|c|}
\hline
\multicolumn{1}{|c|}{name} & unit & formula\tabularnewline
\hline
\hline
Biot coefficient & - & $\alpha_{B}\left(x\right)=1-\frac{K_{d}\left(x\right)}{K_{s}}$\tabularnewline
\hline
storage coefficient & ${\color{revisionpurple}\mathrm{Pa}^{-1}}$ & $S\left(x\right)=\frac{\alpha_{B}\left(x\right)-\phi\left(x\right)}{K_{s}}+\frac{\phi\left(x\right)}{K_{f}}$\tabularnewline
\hline
Lamé's first parameter & Pa & $\lambda\left(x\right)=K_{d}\left(x\right)-\frac{2}{3}\mu\left(x\right)$\tabularnewline
\hline
\end{tabular}
\par\end{centering}
\caption{Derived material properties entering the PDEs; $x$ is the spatial
variable}\label{tab:Derived-material-properties}
\end{table}

In contrast to previously used models \cite{rutqvist_modeling_2009,Beresova2022Bayesian,Luber2024},
we model hydraulic conductivity and porosity using non-stationary
Gaussian random fields (GRFs). The covariance function is
\begin{equation}
C\left(x_{i},x_{j}\right)=\sigma\left(x_{i}\right)\sigma\left(x_{j}\right)\frac{2\ell\left(x_{i}\right)\ell\left(x_{j}\right)}{\ell\left(x_{i}\right)^{2}+\ell\left(x_{j}\right)^{2}}\exp\left(-\frac{\left\Vert x_{i}-x_{j}\right\Vert ^{2}}{\ell\left(x_{i}\right)^{2}+\ell\left(x_{j}\right)^{2}}\right),\label{eq:Covariance_Paciorek_Schervish-1}
\end{equation}
{\color{revisionpurple}which is the non-stationary covariance construction from \cite{paciorek_spatial_2006},} where
$x_{i}$, $x_{j}$ are spatial coordinates, $\ell\left(x\right)$
is the spatially varying correlation length, and $\sigma\left(x\right)$
is the standard deviation.{\color{revisionpurple}~Here, the random fields are represented on the finite-element mesh. The covariance matrix \(\mathbf{C}\in\mathbb{R}^{N\times N}\) has entries \(\mathbf{C}_{ij}=C(x_i,x_j)\), where \(x_i\) and \(x_j\) are mesh nodes. Denote \(\lambda_m\) and \(\mathbf{v}_m\)  its eigenpairs, and \(\xi_m\sim\mathcal{N}(0,1)\). The truncated Karhunen-Loeve (KL) expansion is then given in the nodal form as \[g_M(x_i)=\sum_{m=1}^{M}\xi_m\sqrt{\lambda_m}\,\mathbf{v}_m(i),\qquad M\leq N\] for a mesh node \(x_i\). }

\begin{figure}[H]
\begin{centering}
\subfloat[$\ell\left(x\right)$ for $K\left(x\right)$ and $\phi\left(x\right)$]{\includegraphics[width=0.46\textwidth]{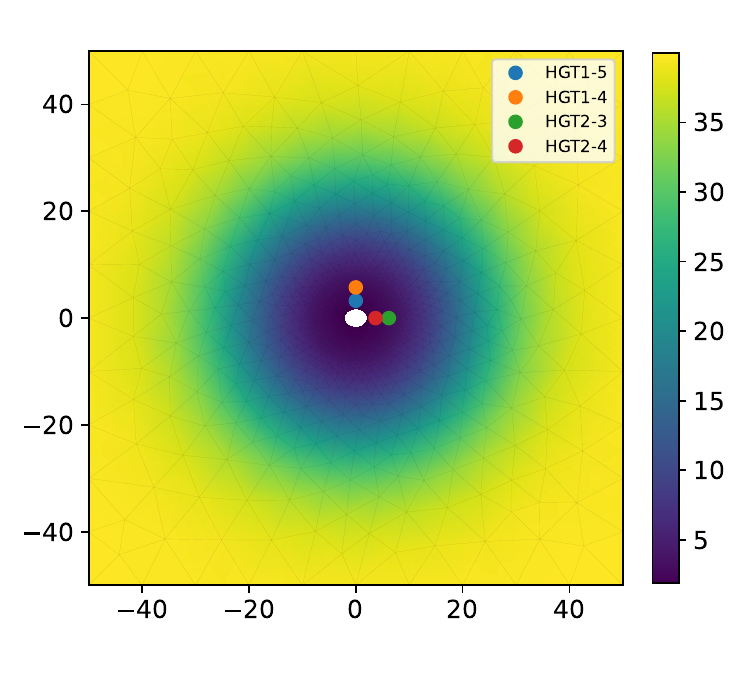}}\subfloat[$\sigma_{K}\left(x\right)$ for $K\left(x\right)$]{\includegraphics[width=0.46\textwidth]{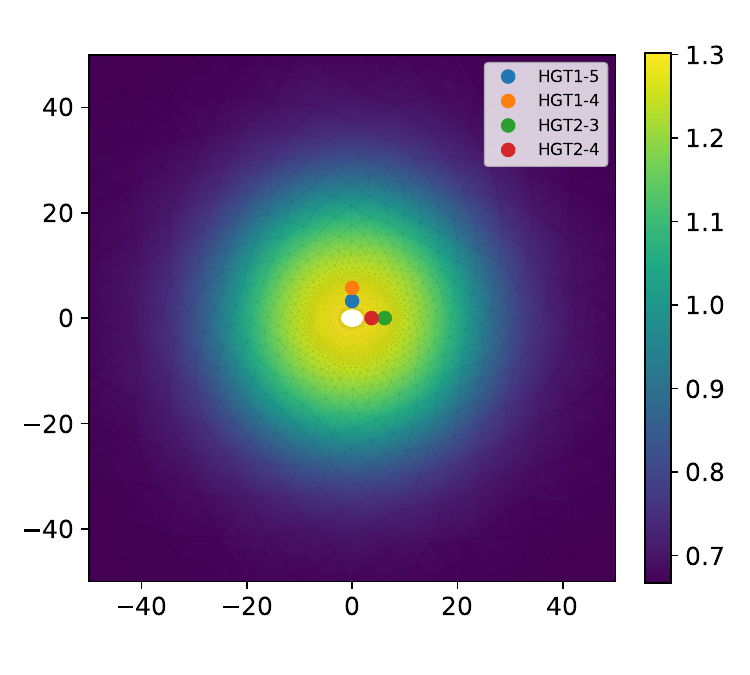}}\caption{$\ell\left(x\right)$ field used both for hydraulic conductivity
and porosity; $\sigma_{K}\left(x\right)$ field used for hydraulic
conductivity (for porosity, constant $\sigma_{\phi}\left(x\right)=0.5$
was chosen)}\label{fig:-fields-ell-sigma}
\par\end{centering}
\end{figure}

\begin{figure}[H]
\centering{}\subfloat[$M_{K}=N=1884$]{\includegraphics[width=0.46\textwidth]{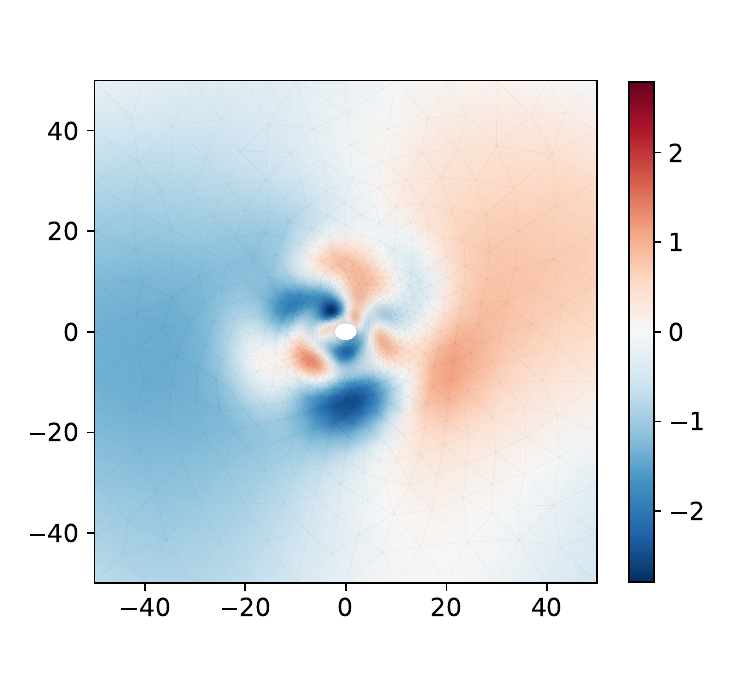}}\subfloat[$M_{K}=40$]{\includegraphics[width=0.46\textwidth]{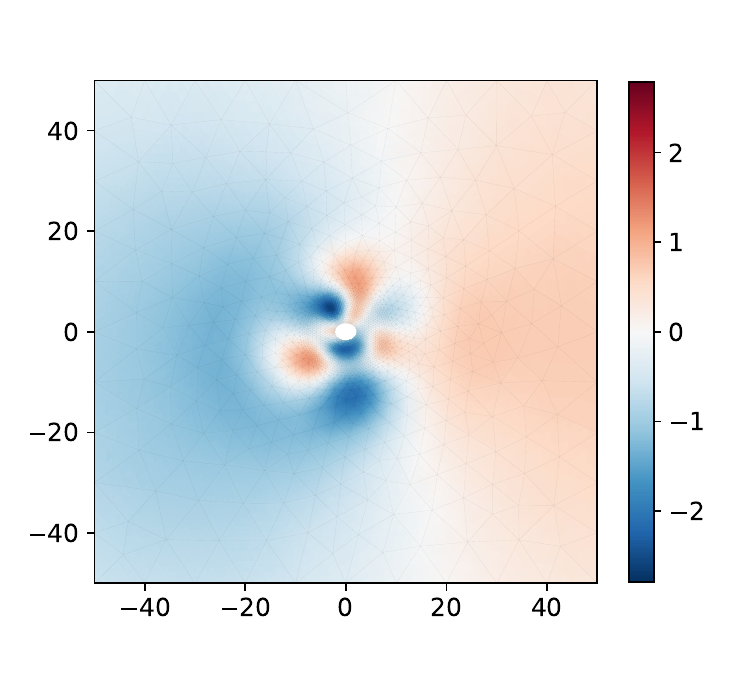}}\caption{Random realization of $\text{GRF}^{M_{K}}_{K}\left(x\right)$ before
and after truncation}\label{fig:random_GRF_realization_Kh}
\end{figure}

\begin{figure}[H]
\centering{}\subfloat[$M_{\phi}=N=1884$]{\includegraphics[width=0.46\textwidth]{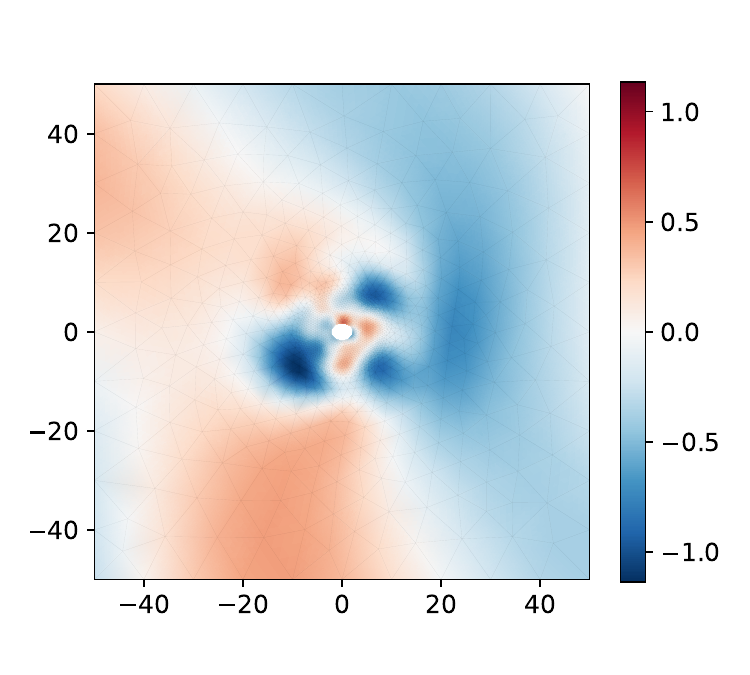}}\subfloat[$M_{\phi}=40$]{\includegraphics[width=0.46\textwidth]{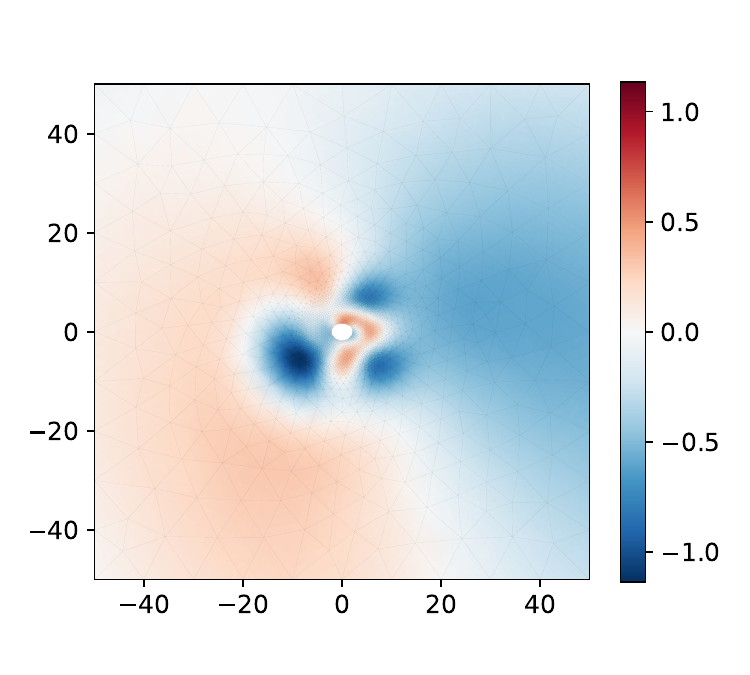}}\caption{Random realization of $\text{GRF}^{M_{\phi}}_{\phi}\left(x\right)$
before and after truncation}\label{fig:random_GRF_realization_por}
\end{figure}

The functions $\ell\left(x\right)$ and $\sigma_{K}\left(x\right)$
and $\sigma_{\phi}\left(x\right)$
 for hydraulic conductivity and porosity are defined as functions
of the distance from the tunnel boundary, see Fig.~\ref{fig:-fields-ell-sigma}.
This yields truncated random fields $\text{GRF}^{M_{K}}_{K}\left(x\right)$
and $\text{GRF}^{M_{\phi}}_{\phi}\left(x\right)$ for $K$ and $\phi$.
Figures \ref{fig:random_GRF_realization_Kh} and \ref{fig:random_GRF_realization_por}
show random realizations of the GRFs and their truncated variants.
The material fields $K\left(x\right)$ and $\phi\left(x\right)$ are
then defined as
\begin{equation}
K\left(x\right)=10^{\text{GRF}^{M_{K}}_{K}\left(x\right)-13},\label{eq:field_Kh}
\end{equation}
\begin{equation}
\phi\left(x\right)=10^{\text{GRF}^{M_{\phi}}_{\phi}\left(x\right)-2.5}.\label{eq:field_por}
\end{equation}
These random field representations define the prior distributions,
i.e. $K\left(x\right)$ and $\phi\left(x\right)$ are represented
using $M_{K}$ and $M_{\phi}$ parameters.

\begin{figure}[H]
\centering{}\subfloat[$K_{d}\left(x\right)$ prior mean]{\includegraphics[width=0.46\textwidth]{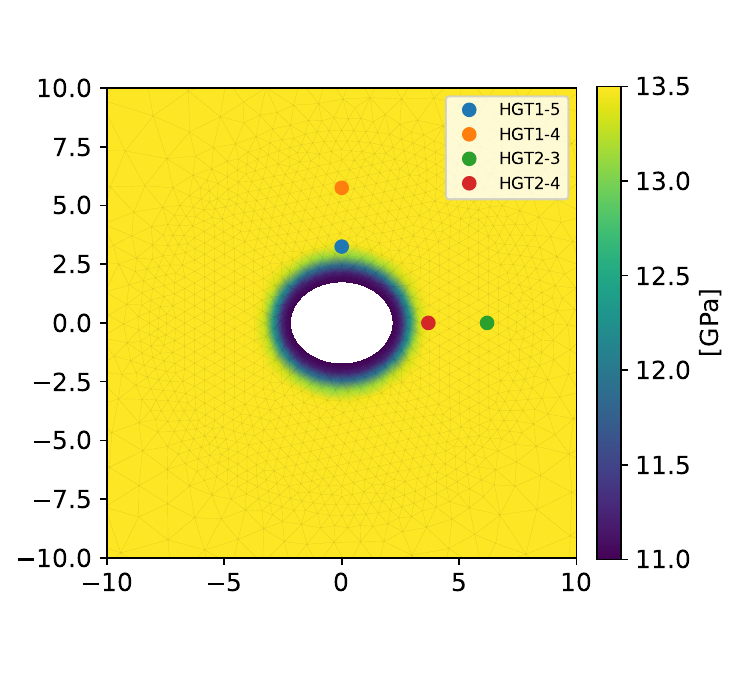}}\subfloat[$\mu\left(x\right)$ prior mean]{\includegraphics[width=0.46\textwidth]{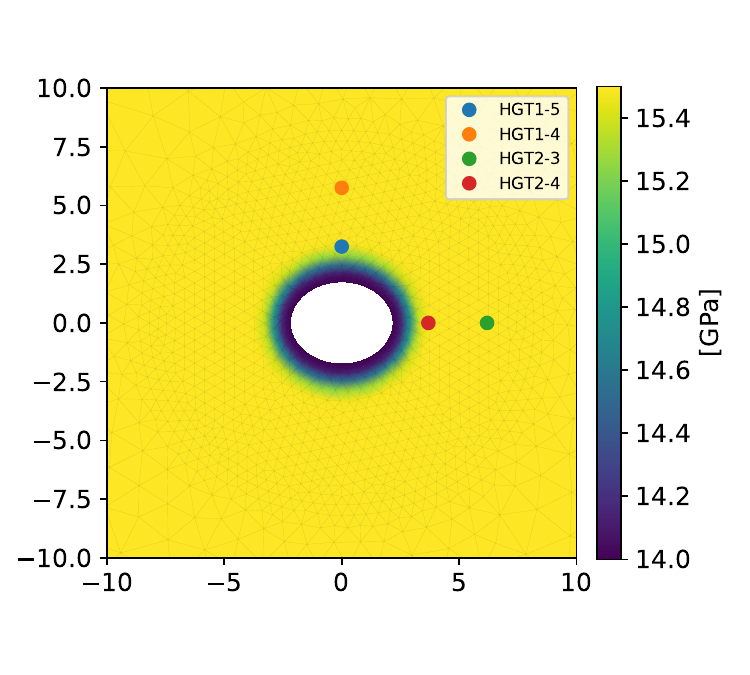}}\caption{Prior mean fields of drained bulk modulus $K_{d}\left(x\right)$
and  $\mu\left(x\right)$}\label{fig:-fields-Kd-shear}
\end{figure}

It remains to specify prior distributions for $K_{d}$, $\mu$, and
$K_{s}$. $K_{s}$ is assumed to be constant with prior distribution
$\mathcal{N}\left(47,2^{2}\right)$ in GPa. $K_{d}$ and $\mu$ are
expected to be lower in the excavation damaged zone (EDZ) and higher
in the intact medium. Therefore, they are represented using two parameters
each: one at the tunnel boundary and one at the outer boundary, with
a smooth transition between them, see Fig.~\ref{fig:-fields-Kd-shear}.
The prior distributions of these parameters are
\begin{equation}
K^{\text{inner}}_{d}\sim\mathcal{N}\left(11,{\color{revisionpurple}1^{2}}\right),\,K^{\text{outer}}_{d}\sim\mathcal{N}\left(13.5,{\color{revisionpurple}1^{2}}\right),\label{prior_Kd}
\end{equation}
\begin{equation}
\mu^{\text{inner}}\sim\mathcal{N}\left(14,{\color{revisionpurple}1^{2}}\right),\,\mu^{\text{outer}}\sim\mathcal{N}\left(15.5,{\color{revisionpurple}1^{2}}\right)\label{prior_shear}
\end{equation}
in GPa.

This defines all input parameters of the forward model and the resulting
prior distribution, consisting of $M_{K}+M_{\phi}+5$ independent
components. The choice of the truncation dimensions $M_{K}$ and $M_{\phi}$
is discussed in the following subsection. Consequently, the forward
map used in the Bayesian inverse problem has the form $G:\mathbb{R}^{M_{K}+M_{\phi}+5}\rightarrow\mathbb{R}^{72}$.
At this point, all components of the Bayesian inverse problem are
defined: $y$, $G$, likelihood, prior, and the posterior distribution
given by the formula (\ref{eq:posterior-1}).

\subsection{Choice of truncation dimensions}

For a Gaussian random field, the first preliminary choice of truncation
dimension can be based on the cumulative variance captured by the
retained KL modes. If $\mathbf{C}$ has eigenvalues $\lambda_{1}\geq\lambda_{2}\geq\cdots$,
then the total prior variance of the GRF is given by the sum of all
eigenvalues, so the fraction of variance captured by the first $M$
modes is
\[
\mathinner{\color{revisionpurple}\frac{\sum^{M}_{m=1}\lambda_{m}}{\sum^{N}_{m=1}\lambda_{m}}}{\color{revisionpurple}.}
\]
 Using this criterion on $\text{GRF}^{M_{K}}_{K}\left(x\right)$
and $\text{GRF}^{M_{\phi}}_{\phi}\left(x\right)$ gives the following
thresholds: For the hydraulic conductivity field, 30 modes capture
approximately 95\% of variance and 53 modes 99\% of variance. For
the porosity field, 32 modes capture approximately 95\% of variance
and 56 modes 99\%.

\begin{figure}[H]
\begin{centering}
\subfloat[Hydraulic conductivity parameters]{\includegraphics[width=0.46\textwidth]{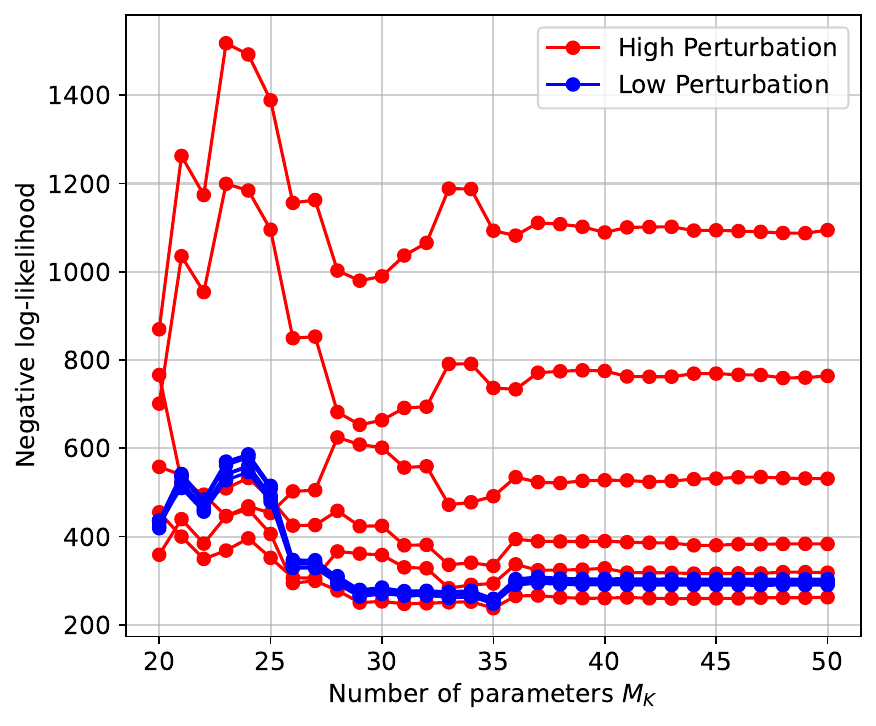}

}\subfloat[Porosity parameters]{\includegraphics[width=0.46\textwidth]{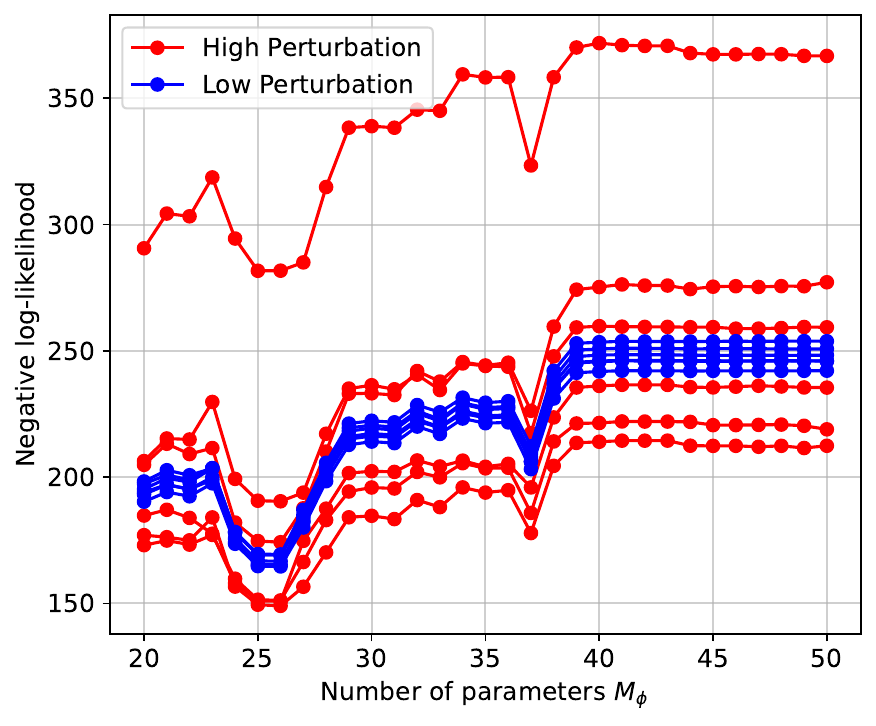}}
\par\end{centering}
\caption{Likelihood-informed truncation study}\label{fig:Truncation-study}
\end{figure}

{\color{revisionpurple}For the final decision, we also considered a likelihood-informed criterion. A parameter vector with the largest posterior density among the samples found so far was used as a \emph{best posterior sample}. Perturbations around this sample were then generated using \(\mathcal{N}(0,0.01^2)\) and \(\mathcal{N}(0,0.1^2)\), and the resulting likelihood responses were compared under different truncation levels. Hydraulic-conductivity parameters and porosity parameters were perturbed separately; see Fig.~\ref{fig:Truncation-study}. Based on this study, \(M_K=M_\phi=40\) was chosen as a sufficient number of modes. Therefore, in what follows, we work with the forward map \(G:\mathbb{R}^{85}\rightarrow\mathbb{R}^{72}\).}

\section{Sampling framework}\label{sec:Sampling-framework}

This section introduces the main sampling ingredients and then combines
them in the delayed-acceptance Hamiltonian Monte Carlo (DAHMC) algorithm
with subchains.

\subsection{Metropolis-Hastings}

We begin with the basic Metropolis-Hastings (MH) algorithm, since
all later methods in this section build on it.  Let $f:\mathbb{R}^{n}\rightarrow\mathbb{R}$
denote the unnormalized posterior density, and let $q\left(\cdot|u\right)$
be a proposal density. Given the current state $u^{\left(k\right)}$,
the MH algorithm (Alg.~\ref{alg:Metropolis-Hastings-algorithm})
draws a proposal $v$ from $q\left(\cdot|u^{\left(k\right)}\right)$
and accepts it with probability
\[
a\left(u^{\left(k\right)},v\right)=\min\!\left(1,\frac{q\left(u^{\left(k\right)}|v\right)f\left(v\right)}{q\left(v|u^{\left(k\right)}\right)f\left(u^{\left(k\right)}\right)}\right).
\]
If the proposal is accepted, then $u^{\left(k+1\right)}=v$; otherwise,
$u^{\left(k+1\right)}=u^{\left(k\right)}$. This acceptance rule ensures
detailed balance with respect to the target distribution and therefore
preserves the posterior distribution as the invariant distribution
of the resulting Markov chain.\textcolor{revisionpurple}{~When the proposal density is symmetric, the ratio of proposal densities cancels and the method reduces to the original Metropolis algorithm \cite{metropolis_equation_1953}; the general Metropolis-Hastings form follows \cite{hastings_monte_1970}.}

The efficiency of the MH algorithm strongly depends on the choice
of the proposal density. If proposals are too conservative (e.g. random
walk with small variance), the chain mixes slowly because of high
autocorrelation. If they are too aggressive (e.g. random walk with
high variance), the acceptance rate becomes too low. In inverse problems
driven by computationally intensive forward models, the main drawback
is that each proposal (either accepted or rejected) requires an evaluation
of the full forward model. This motivates the use of the delayed-acceptance
principle reviewed next.

\begin{algorithm}[H]
\begin{algorithmic}[1]
\Require target density \(f:\mathbb{R}^{n}\to\mathbb{R}\), proposal density \(q\), initial state \(u^{(0)}\) with \(f(u^{(0)})>0\).
\For{\(k=0,1,\ldots\)}
	\State Draw \(v\sim q(\cdot\mid u^{(k)})\).
	\State Accept the proposal with the acceptance probability
	\Statex \hspace{\algorithmicindent}\hspace{\algorithmicindent}\(
		\begin{aligned}
        a(u^{(k)},v)
        &= \min\!\left(
            1,\,
            \frac{q(u^{(k)}\mid v)\,f(v)}
                 {q(v\mid u^{(k)})\,f(u^{(k)})}
        \right).
		\end{aligned}
	\)
	\If{the proposal is accepted }
        \State Set \(u^{(k+1)}=v\).
	\Else
        \State Set \(u^{(k+1)}=u^{(k)}\).
	\EndIf
\EndFor
\end{algorithmic}

\caption{Metropolis-Hastings algorithm}\label{alg:Metropolis-Hastings-algorithm}
\end{algorithm}

\subsection{Delayed acceptance Metropolis-Hastings}

When the dominant computational cost arises from evaluating the forward
model, it is natural to introduce a cheap approximation of the target
density. Delayed acceptance Metropolis-Hastings (DAMH), introduced
by Christen and Fox \cite{christen_markov_2005}, uses such an approximation
to reject unpromising proposals before the expensive forward model
is evaluated.

Let $\widetilde{f}:\mathbb{R}^{n}\rightarrow\mathbb{R}$ denote an
approximate unnormalized posterior density such that $\text{supp}\widetilde{f}\supset\text{supp}f$,
for details on DAMH requirements see \cite{Beresova2022Bayesian}.
Given the current state $u^{\left(k\right)}$ and a proposal $v\sim q\left(\cdot|u^{\left(k\right)}\right)$,
the first stage accepts the proposal with probability
\[
a_{Q,\widetilde{\mu}}\left(u^{\left(k\right)},v\right)=\min\!\left(1,\frac{q\left(u^{\left(k\right)}|v\right)\widetilde{f}\left(v\right)}{q\left(v|u^{\left(k\right)}\right)\widetilde{f}\left(u^{\left(k\right)}\right)}\right).
\]
The exact density $f\left(v\right)$ is evaluated only if the proposal
passes this preliminary test. The second-stage acceptance probability
is then
\[
a_{\widetilde{Q},\mu}\left(u^{\left(k\right)},v\right)=\min\!\left(1,\frac{\widetilde{f}\left(u^{\left(k\right)}\right)f\left(v\right)}{\widetilde{f}\left(v\right)f\left(u^{\left(k\right)}\right)}\right).
\]
This two-stage construction preserves the correct posterior distribution
while reducing the number of expensive evaluations of the full forward
model, see Alg.~\ref{alg:Delayed-acceptance-Metropolis-Ha}. In the
present paper, the approximation $\widetilde{f}$ is induced by a
surrogate approximation of the forward map $G$.

\begin{algorithm}[H]
\begin{algorithmic}[1]
\Require exact density \(f\), surrogate density \(\widetilde f\), proposal density \(q\), initial state \(u^{(0)}\)
\State with \(f(u^{(0)})>0\).
\For{\(k=0,1,\ldots\)}
	\State Draw \(v\sim q(\cdot\mid u^{(k)})\).
	\State Accept the proposal with the first-stage acceptance probability
	\Statex \hspace{\algorithmicindent}\hspace{\algorithmicindent}\(
		\begin{aligned}
		a_{Q,\widetilde\mu}(u^{(k)},v)
		&= \min\!\left(
			1,\,
			\frac{q(u^{(k)}\mid v)\,\widetilde f(v)}
				{q(v\mid u^{(k)})\,\widetilde f(u^{(k)})}
		\right).
		\end{aligned}
	\)
	\If{the first stage accepts}
		\State Evaluate the exact density \(f(v)\).
		\State Accept the proposal with the second-stage acceptance probability
		\Statex \hspace{\algorithmicindent}\hspace{\algorithmicindent}\(
		\begin{aligned}
		a_{\widetilde Q,\mu}(u^{(k)},v)
		&= \min\!\left(
            1,\,
            \frac{\widetilde f(u^{(k)})\,f(v)}
                 {\widetilde f(v)\,f(u^{(k)})}
        \right).
		\end{aligned}
		\)
		\If{the second stage accepts}
		\State Set \(u^{(k+1)}=v\).
		\Else
		\State Set \(u^{(k+1)}=u^{(k)}\).
		\EndIf
	\Else
		\State Set \(u^{(k+1)}=u^{(k)}\).
	\EndIf
\EndFor
\end{algorithmic}

\caption{Delayed acceptance Metropolis-Hastings algorithm}\label{alg:Delayed-acceptance-Metropolis-Ha}
\end{algorithm}

\subsection{pCN proposal}

The preconditioned Crank-Nicolson (pCN) proposal \cite{cotter_mcmc_2013}
is a dimension-robust alternative to random-walk proposals. In this
work, it serves as a baseline for comparison, not as a main ingredient
of the proposed sampling framework. For a Gaussian prior, pCN is reversible
with respect to the prior. If $u$ denotes the current state and $\xi\sim\mathcal{N}(0,C_{0})$
an independent Gaussian prior draw, then the pCN proposal is
\[
v=\sqrt{1-\beta^{2}}\,u+\beta\,\xi,\qquad\beta\in(0,1].
\]

Compared with random-walk proposals, pCN has one main tuning parameter
$\beta$. This is convenient in practice; however, the choice of $\beta$
still strongly influences acceptance and mixing.

\subsection{Hamiltonian proposal}

The efficiency of MH-based algorithms strongly depends on the proposal
strategy. For high-dimensional and strongly correlated posterior distributions,
random-walk proposals often lead to slow exploration of the state
space due to high autocorrelation of the resulting chain. Improving
mixing has two main aspects:
\begin{enumerate}
\item The proposed sample should be ``distant'' from the original sample
to reduce correlation.
\item The acceptance probability should be high, so the chain has a low
probability of staying in the original position.
\end{enumerate}
To address both objectives, we consider the Hamiltonian proposal
according to \cite{neal_mcmc_2011}. The key idea is to augment the
parameter vector by an auxiliary momentum vector and to generate proposals
by approximately following a Hamiltonian trajectory. Long Hamiltonian
trajectories allow distant proposals, which is the first aspect of
the decorrelation goal.

Let $u\in\mathbb{R}^{n}$ denote the parameter vector and let $p\in\mathbb{R}^{n}$
be an auxiliary momentum variable with distribution $\mathcal{N}\left(0,M\right)$.\textcolor{revisionpurple}{~Here, \(M\) is a symmetric positive-definite mass matrix.}
Define the potential energy $\mathcal{U}\left(u\right)=-\log f\left(u\right)$
up to an additive constant, the kinetic energy $\mathcal{K}\left(p\right)=\frac{1}{2}p^{T}M^{-1}p$,
and the Hamiltonian
\[
\mathcal{H}\left(u,p\right)=\mathcal{U}\left(u\right)+\mathcal{K}\left(p\right).
\]
 The continuous Hamiltonian dynamics are given by
\[
\frac{\mathrm{d}u}{\mathrm{d}t}=M^{-1}p,\qquad\frac{\mathrm{d}p}{\mathrm{d}t}=\mathbin{\color{revisionpurple}-}{\color{revisionpurple}\nabla\mathcal{U}}\mathinner{\color{revisionpurple}\left(u\right)}.
\]
 Their exact flow preserves the Hamiltonian and is volume preserving,
which makes it suitable for constructing proposals in the Metropolis-Hastings
algorithm.

In computations, the continuous dynamics are replaced by a reversible
and volume-preserving symplectic integrator, typically the leapfrog
scheme. Starting from the current state $u^{\left(k\right)}$, one
first draws a fresh momentum $p^{\left(0\right)}\sim\mathcal{N}\left(0,M\right)$
and then performs $L$ leapfrog steps with step size $\varepsilon>0$:
\[
p^{\left(\ell+\frac{1}{2}\right)}=p^{\left(\ell\right)}\mathbin{\color{revisionpurple}-}{\color{revisionpurple}\frac{\varepsilon}{2}\nabla\mathcal{U}}\mathinner{\color{revisionpurple}\left(u^{\left(\ell\right)}\right)},
\]
 {}
\[
u^{\left(\ell+1\right)}=u^{\left(\ell\right)}+\varepsilon M^{-1}p^{\left(\ell+\frac{1}{2}\right)},
\]
 {}
\[
p^{\left(\ell+1\right)}=p^{\left(\ell+\frac{1}{2}\right)}\mathbin{\color{revisionpurple}-}{\color{revisionpurple}\frac{\varepsilon}{2}\nabla\mathcal{U}}\mathinner{\color{revisionpurple}\left(u^{\left(\ell+1\right)}\right)},\qquad\ell=0,\ldots,L-1.
\]
{\color{revisionpurple}Then, \((v,-p^{(L)})\) is taken as the proposal. The momentum flip makes the numerical Hamiltonian map reversible, while \(\mathcal{K}(p)=\mathcal{K}(-p)\), so the acceptance probability is numerically unchanged.}
The resulting proposal is accepted with probability
\begin{equation}
a_{\mathrm{HMC}}\left(\left(u^{\left(k\right)},p^{\left(0\right)}\right),\mathinner{\color{revisionpurple}\left(v,-p^{\left(L\right)}\right)}\right)=\min\!\left(1,\exp\left(\mathbin{\color{revisionpurple}-}{\color{revisionpurple}\mathcal{H}\left(v,-p^{\left(L\right)}\right)+\mathcal{H}}\mathinner{\color{revisionpurple}\left(u^{\left(k\right)},p^{\left(0\right)}\right)}\right)\right).\label{HMC_acceptance}
\end{equation}
 If the numerical integration were exact, this acceptance probability
would equal one. In practice, the correction compensates for the discretization
error introduced by the leapfrog approximation but remains close to
1, which is the second aspect of the decorrelation goal.

In the proposed sampling framework, Hamiltonian trajectories are not
used directly with the full forward model. Instead, they are generated
using a surrogate-based approximation of the target and serve as an
efficient first-stage proposal mechanism inside the delayed-acceptance
framework.\textcolor{revisionpurple}{~With a fixed surrogate model, this is an ordinary delayed-acceptance proposal and the second-stage correction preserves the exact posterior. Surrogate updates, if used, are confined to burn-in.}
The discussion so far is finite-dimensional, with the unknown represented
by a vector in $\mathbb{R}^{n}$ . A dimension-robust Hamiltonian
proposal can also be defined in function space, similarly to pCN,
see \cite{beskos_hybrid_2011}.

\subsection{Subchain of Hamiltonian proposals}

The delayed-acceptance framework can be generalized by replacing the
single first-stage proposal with a short subchain targeting the approximate
posterior distribution. This perspective is closely related to \cite{lykkegaard_multilevel_2023}.
Let $\widetilde{\mu}$ denote the approximate posterior measure with
density proportional to $\widetilde{f}$, and let $\widetilde{K}$
be a Markov transition kernel reversible with respect to $\widetilde{\mu}$.
Starting from the current outer state $u^{\left(k\right)}$, define
an inner chain by $w^{\left(0\right)}=u^{\left(k\right)}$ and generate
$w^{\left(1\right)},\ldots,w^{\left(J\right)}$ using the kernel $\widetilde{K}$.
The endpoint $v=w^{\left(J\right)}$ is then used as the first-stage
proposal for the exact target. In other words, the usual proposal
density is replaced by the $J$-step kernel $\widetilde{K}^{J}$.

This replacement leads to a particularly simple acceptance formula.
Since $\widetilde{K}$ is reversible with respect to $\widetilde{\mu}$,
the same is true for $\widetilde{K}^{J}$, that is,
\[
\widetilde{f}\left(u\right)\widetilde{K}^{J}\left(u,\mathrm{d}v\right)=\widetilde{f}\left(v\right)\widetilde{K}^{J}\left(v,\mathrm{d}u\right).
\]
 Therefore, if the outer chain targeting the exact density $f$ uses
$\widetilde{K}^{J}$ as its proposal kernel, then the second-stage
acceptance probability has the standard form
\[
a_{\mathrm{DA}}\left(u,v\right)=\min\!\left(1,\frac{\widetilde{f}\left(u\right)f\left(v\right)}{\widetilde{f}\left(v\right)f\left(u\right)}\right).
\]
 The detailed balance condition for the outer chain follows immediately
from the reversibility of $\widetilde{K}^{J}$:
\[
f\left(u\right)a_{\mathrm{DA}}\left(u,v\right)\widetilde{K}^{J}\left(u,\mathrm{d}v\right)=f\left(v\right)a_{\mathrm{DA}}\left(v,u\right)\widetilde{K}^{J}\left(v,\mathrm{d}u\right).
\]
 Hence, the endpoint of a finite subchain targeting the surrogate
posterior can be used as a valid proposal for the exact posterior.

In the general discussion above, \textcolor{revisionpurple}{ \(\widetilde K\) may be any reversible kernel targeting the approximate posterior \(\widetilde\mu\).}
Here, we consider an MH kernel with a Hamiltonian proposal targeting
$\widetilde{\mu}$.

\subsection{DAHMC algorithm with subchains}\label{subsec:DAHMC-algorithm-with}

Combining the previous ingredients leads to the DAHMC algorithm with
subchains. In each outer iteration, a fixed-length Hamiltonian subchain
targeting the surrogate posterior is constructed,  and the endpoint
of that subchain is then corrected by delayed acceptance with respect
to the exact posterior.

\textcolor{revisionpurple}{The surrogate model is fixed within each outer iteration: all surrogate likelihoods, gradients, inner HMC decisions, and delayed-acceptance ratios in one outer iteration use the same fixed surrogate model. Retraining affects only later iterations during burn-in.}

\begin{algorithm}[H]
\begin{algorithmic}[1]
\Require exact density \(f\), initial surrogate density \(\widetilde f^{(0)}\), subchain length \(J\),
\State leapfrog step size \(\varepsilon\), leapfrog step count \(L\), mass matrix \(M\), initial state \(u^{(0)}\).
\For{\(k=0,1,\ldots\)}
    \State Set \(w^{(0)}=u^{(k)}\).
    \For{\(j=0,\ldots,J-1\)}
        \State Draw \(p^{(0)}\sim\mathcal{N}(0,M)\).
        \State Compute \(L\) surrogate-driven leapfrog steps of size \(\varepsilon\) starting from \((w^{(j)},p^{(0)})\)
        \State and ending in \((z,p^{(L)})\).
        \State Accept \(z\) with the Hamiltonian acceptance probability from Eq.~\eqref{HMC_acceptance}.
        \If{\(z\) is accepted}
            \State Set \(w^{(j+1)}=z\).
        \Else
            \State Set \(w^{(j+1)}=w^{(j)}\).
        \EndIf
    \EndFor
    \State Set \(v=w^{(J)}\).
    \State Accept the proposal with the second-stage acceptance probability
    \Statex \hspace{\algorithmicindent}\hspace{\algorithmicindent}~\(
        \begin{aligned}
        a_{\mathrm{DA}}(u^{(k)},v)
        &= \min\!\left(
            1,\,
            \frac{\widetilde f^{(k)}(u^{(k)})\,f(v)}
                 {\widetilde f^{(k)}(v)\,f(u^{(k)})}
        \right).
        \end{aligned}
    \)
    \If{the second stage accepts}
        \State Set \(u^{(k+1)}=v\).
    \Else
        \State Set \(u^{(k+1)}=u^{(k)}\).
    \EndIf
    \State Set \(\widetilde f^{(k+1)}\) either as \(\widetilde f^{(k)}\) or its update.
\EndFor
\end{algorithmic}

\caption{DAHMC algorithm with Hamiltonian proposal subchains}
\end{algorithm}

\subsection{Parallel scheme with surrogate retraining}

\begin{figure}[H]
\includegraphics[width=1\textwidth]{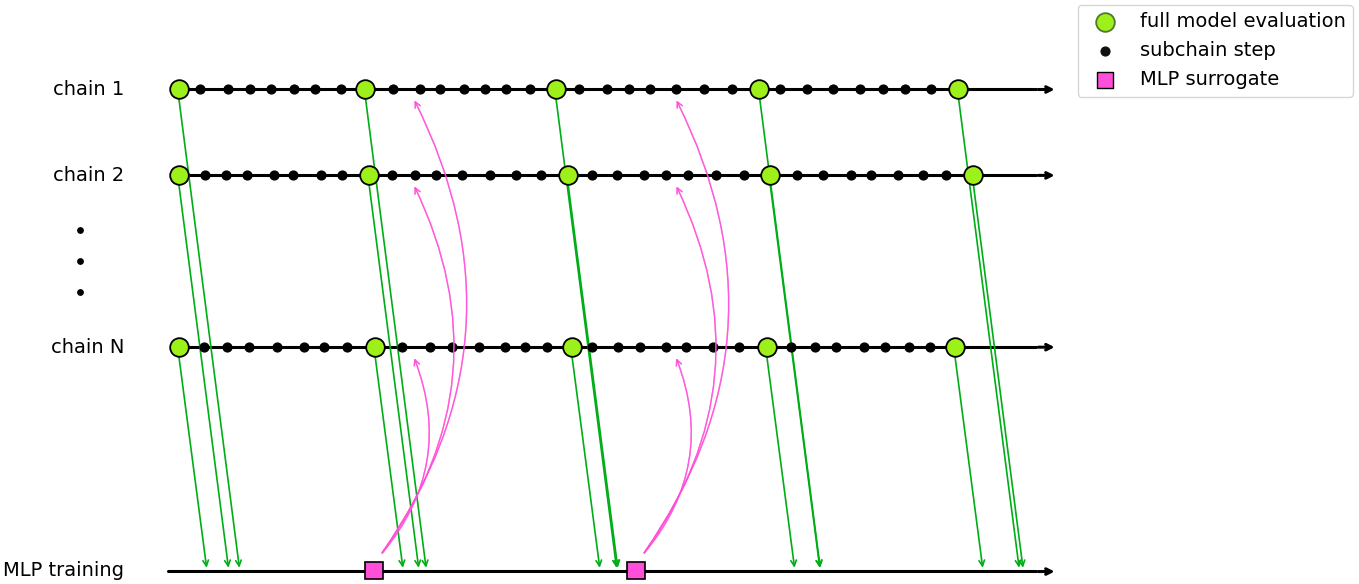}

\caption{Parallel sampling architecture with sampler processes, a shared collector,
and surrogate retraining.}\label{fig:processes_scheme}

\end{figure}

The sampling framework is implemented as a parallel delayed-acceptance
scheme in which multiple sampling processes share one surrogate model.
Each sampler generates its own Markov chain and requests exact forward
model solutions only when needed by the delayed-acceptance correction.
The resulting parameter-observation snapshots are sent to a collector
process, which accumulates the training data and retrains the neural-network
surrogate model. Figure~\ref{fig:processes_scheme} shows a schematic
communication diagram between samplers, the collector, and the surrogate-updater
module. \textcolor{revisionpurple}{The updated surrogate models are distributed to the samplers only during the burn-in phase.}

In the current setting, the surrogate is a fully connected PyTorch
multilayer perceptron (MLP) with three hidden layers of width 256
and SiLU activation functions. The outputs are normalized by a fixed
mean vector and scale vector derived from the reference observations.
 The main advantage of choosing a MLP as a surrogate model is the
simplicity of obtaining derivatives of the approximate forward map
through automatic differentiation. This makes gradient-based proposals
such as Hamiltonian subchains straightforward to implement. Its main
disadvantage is the relatively large number of hyperparameters that
must be tuned, including architecture and training process settings.

\section{Sampling efficiency}\label{sec:Efficiency-analysis}

Sampling efficiency is assessed by how many effectively independent
posterior samples are obtained for a given computational cost. In
practice, we estimate this through the integrated autocorrelation
time (IAT) and the corresponding effective sample size (ESS). Lower
autocorrelation and larger ESS indicate more efficient exploration
of the posterior. The IAT tends to be underestimated when only short
chains are available, so the results in this section should be read
as practical finite-run diagnostics rather than exact asymptotic quantities.

The numerical comparison in this section considers three samplers:
DAHMC, DAMH with a pCN subchain, and MH with a pCN proposal and no
surrogate. The tuning parameters were fixed after burn-in, including
the pCN parameter, the pretrained MLP surrogate, and the HMC mass
matrix.

For DAHMC with subchain length $J$, leapfrog count $L$, full model
cost $c_{F}$, surrogate evaluation cost $c_{E}$, surrogate gradient
cost $c_{J}$, and probability $p_{\mathrm{move}}$ that at least
one inner proposal is accepted, the cost of one outer step is approximated
by
\[
T_{\mathrm{step}}(J,L)\approx J\big((L+1)c_{J}+c_{E}\big)+p_{\mathrm{move}}c_{F}.
\]
For the present TSX benchmark, we take $c_{F}\approx854.7$ ms, $c_{E}\approx0.051$
ms, and $c_{J}\approx0.729$ ms for the pretrained fixed MLP.

The results presented in Section~\ref{sec:Geotechnical-results}
are based on a run with the following settings: $J=25$, $\varepsilon=0.08$,
$L=20$. Since $p_{\mathrm{move}}=1$ in the reported run, the outer
step cost is
\[
T_{\mathrm{step}}(25,20)\approx854.7+25\big(0.729\cdot21+0.051\big)\approx1239\ \mathrm{ms}.
\]
The estimated IAT was $\tau=23$. Therefore, the cost per effective
sample corresponded approximately to the cost of $\frac{\tau}{c_{F}}T_{\mathrm{step}}(25,20)\approx33.4$
evaluations of the full model.

For the pCN subchain run, $\beta=0.07$ and $J=100$ were used, and
the step cost was
\[
T^{pCN}_{\mathrm{step}}(J)\approx Jc_{E}+p_{\mathrm{move}}c_{F}=860\ \mathrm{ms}.
\]
The estimated IAT was $\tau=125$ and $p_{\mathrm{move}}=1$. Therefore,
the cost per effective sample corresponded approximately to the cost
of $\frac{\tau}{c_{F}}T^{pCN}_{\mathrm{step}}(100)\approx125.8$ evaluations
of the full model. Relative to the DAHMC cost above, this corresponds
to a DAHMC speedup factor of about 3.8 over the pCN-subchain variant.

In the case of the pure MH run with pCN proposal and $\beta=0.07$,
a reliable IAT estimate was not obtained because of the limited chain
length. We therefore report only the qualitative conclusion that the
IAT (i.e., the cost per effective sample measured in full-model evaluations)
 was larger than 2000. Relative to the DAHMC cost above, this corresponds
to a DAHMC speedup factor of more than 60 over the pure MH baseline.

\section{Posterior distribution description}\label{sec:Geotechnical-results}

This section contains the analysis of the posterior distribution of
the benchmark problem using samples obtained by MCMC methods. The
forward model, inverse problem formulation, and sampling efficiency
were discussed in previous sections. Here, we focus on the structure
of the posterior itself, the marginal behavior, and the induced uncertainty
in the spatial fields.

{\color{revisionpurple}The parameter space consists of the \(M_K=40\) hydraulic-conductivity KL coefficients, the \(M_\phi=40\) porosity KL coefficients, and the five scalar mechanical parameters.  }The
prior distribution has independent components. Internally, the sampling
framework uses prior components transformed into $\mathcal{N}\left(0,1\right)$.
This affects only the five scalar parameters; the KL coefficient priors
are already standard normal.

\begin{figure}[H]
\begin{centering}
\subfloat[\label{fig:tsx2-inner-correlation}Posterior correlation matrix]{\includegraphics[width=0.46\textwidth]{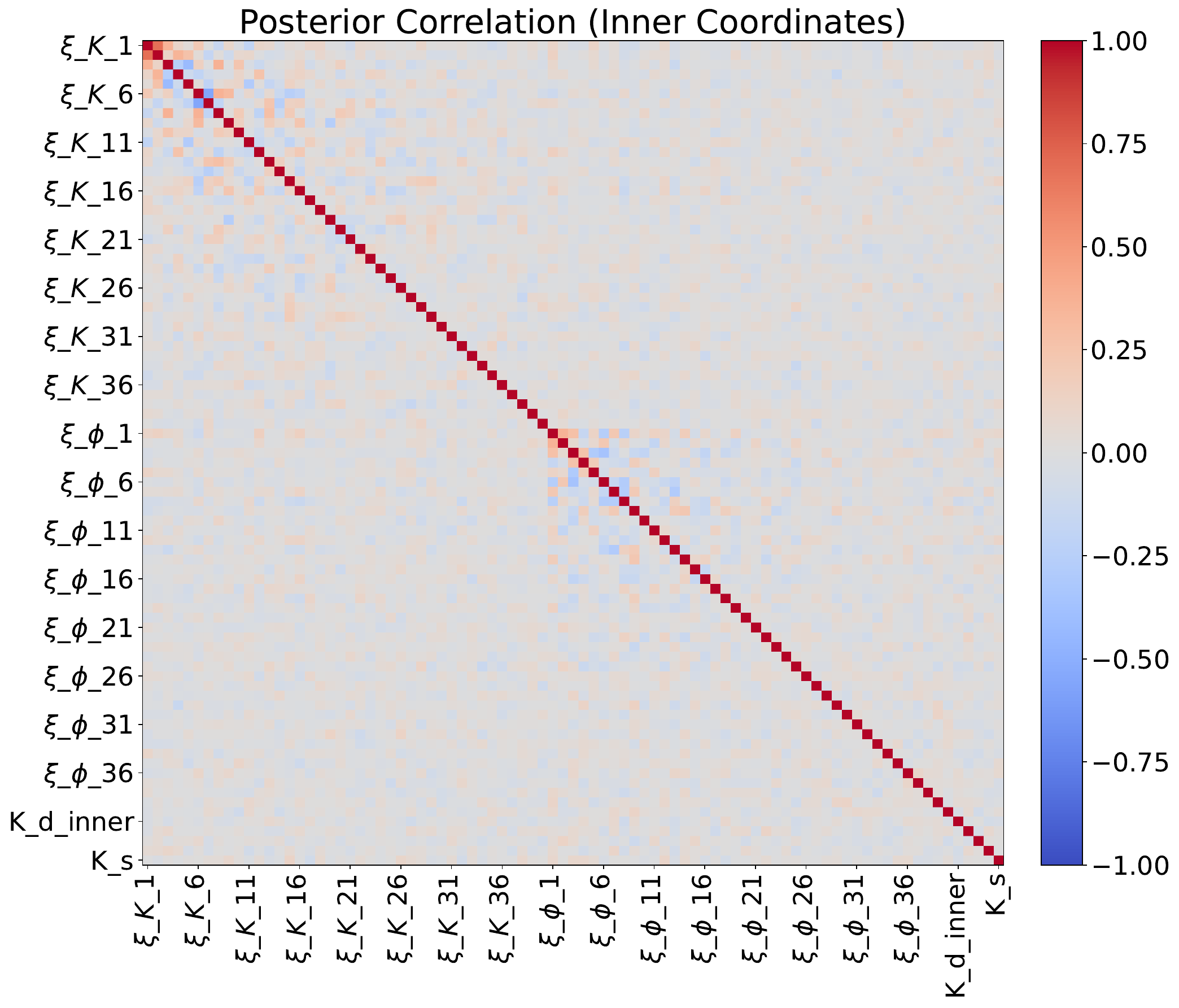}
}\subfloat[\label{fig:tsx2-inner-covariance}Posterior covariance matrix]{\includegraphics[width=0.46\textwidth]{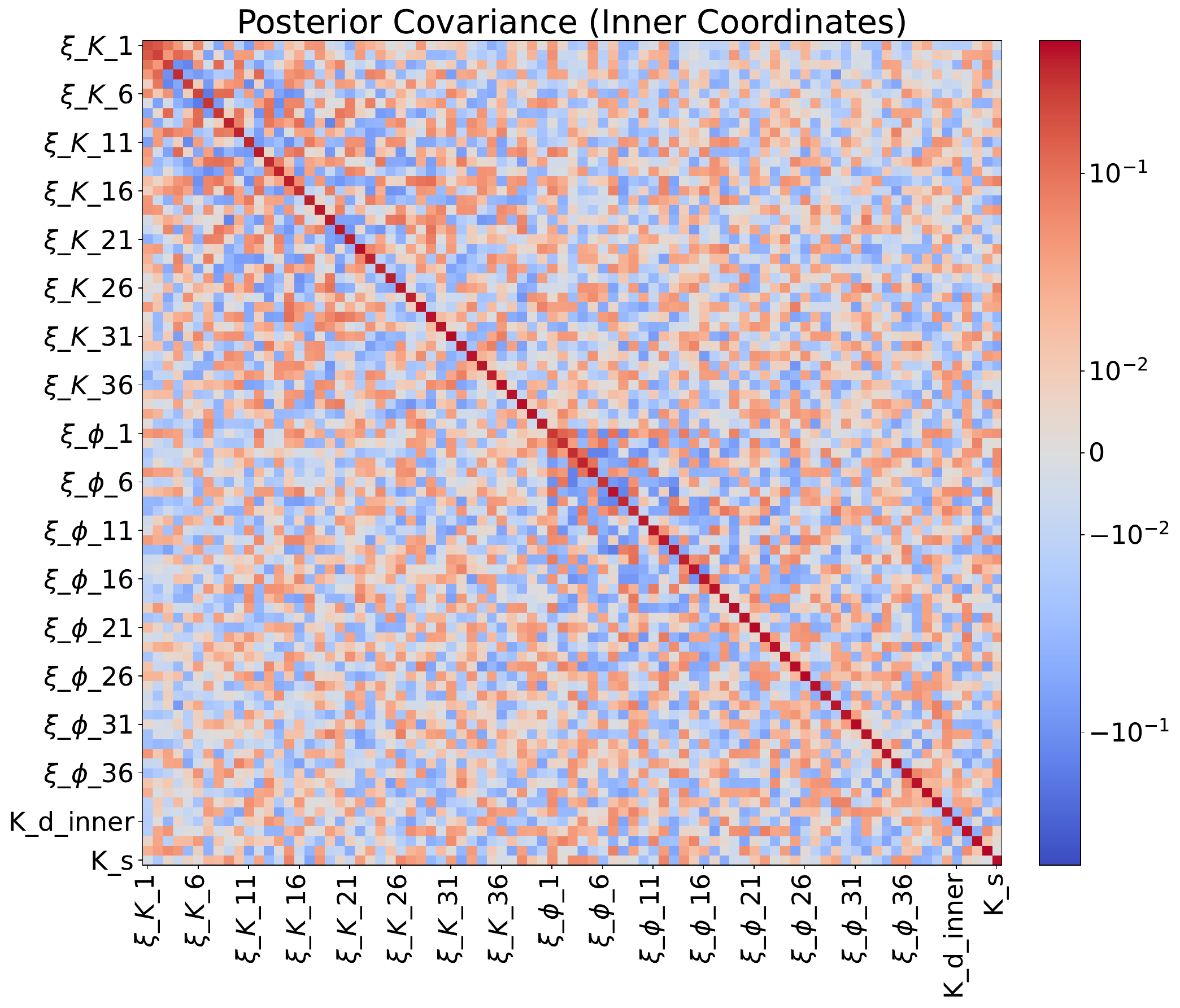}
}
\par\end{centering}
\caption{Posterior correlation and covariance matrix in inner coordinates}

\end{figure}

A basic view is provided by the correlation and covariance matrices
in the internal coordinates; see Figures~\ref{fig:tsx2-inner-correlation}
and \ref{fig:tsx2-inner-covariance}. We can observe couplings among
the KL coefficients and between the field coefficients and the scalar
mechanical parameters.

\begin{figure}[H]
\centering \includegraphics[width=0.9\textwidth]{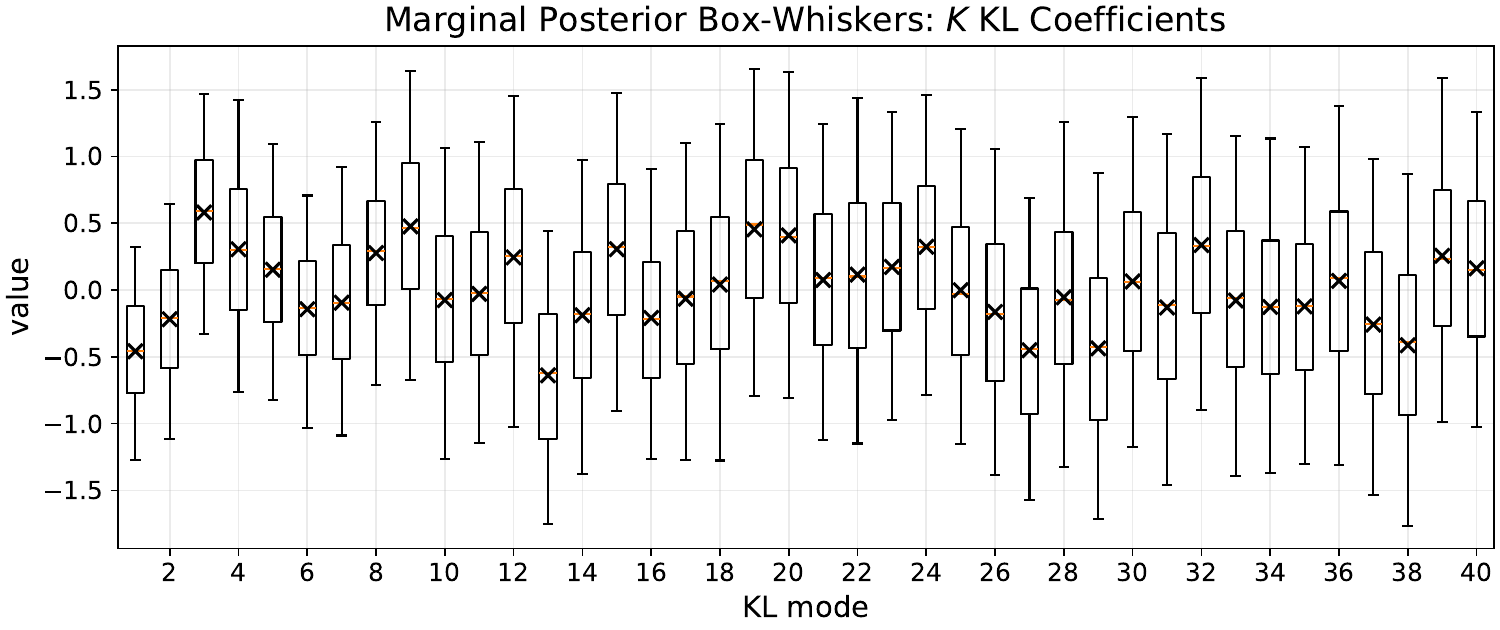}
\caption{Marginal posterior distributions for the hydraulic conductivity KL
coefficients}
\label{fig:tsx2-box-k}
\end{figure}

\begin{figure}[H]
\centering \includegraphics[width=0.9\textwidth]{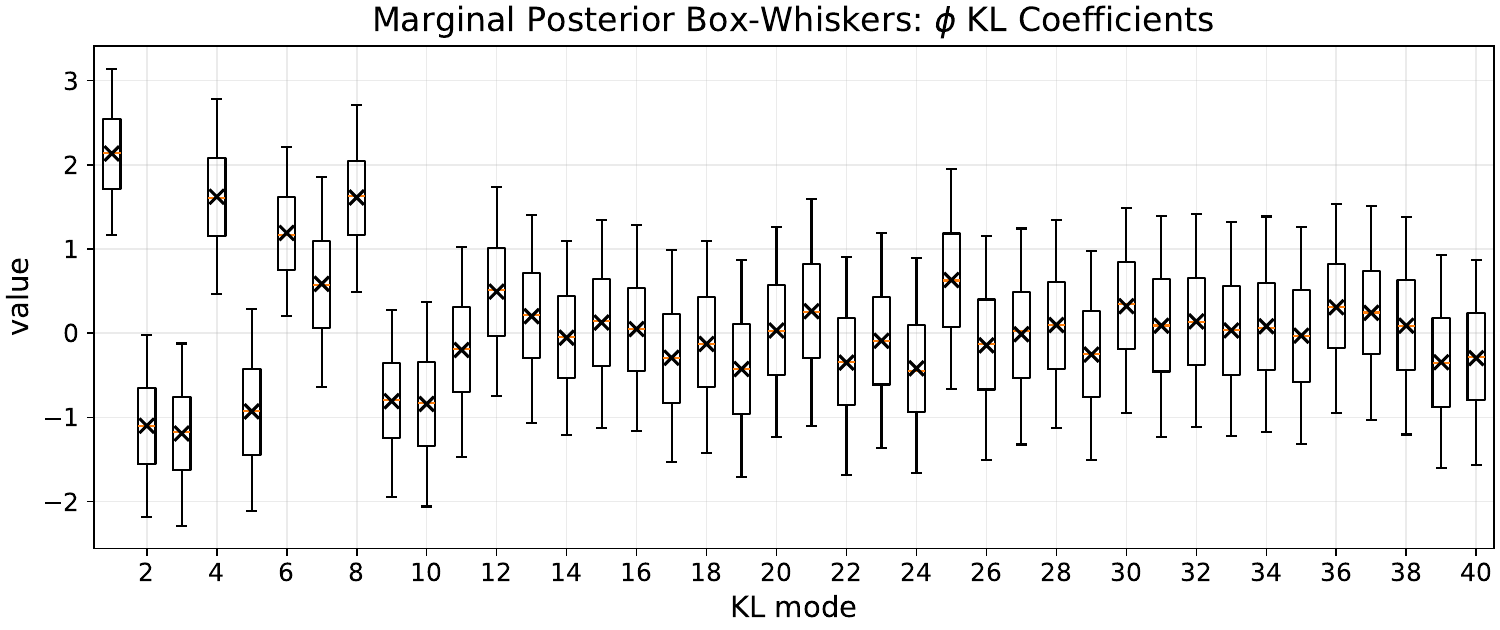}
\caption{Marginal posterior distributions for the porosity KL coefficients}
\label{fig:tsx2-box-phi}
\end{figure}

The KL coefficients associated with $K$ and $\phi$ are summarized
by box--whisker plots in Figures~\ref{fig:tsx2-box-k} and \ref{fig:tsx2-box-phi}.
The boxes show the interquartile range, whiskers indicate the $5\%$
and $95\%$ quantiles, the horizontal line is the median, and the
cross marks the posterior mean. Several leading modes are clearly
shifted away from zero, indicating non-negligible posterior updating
in both fields, compared to the standard Gaussian priors.

\begin{figure}[H]
\centering \includegraphics[width=1\textwidth]{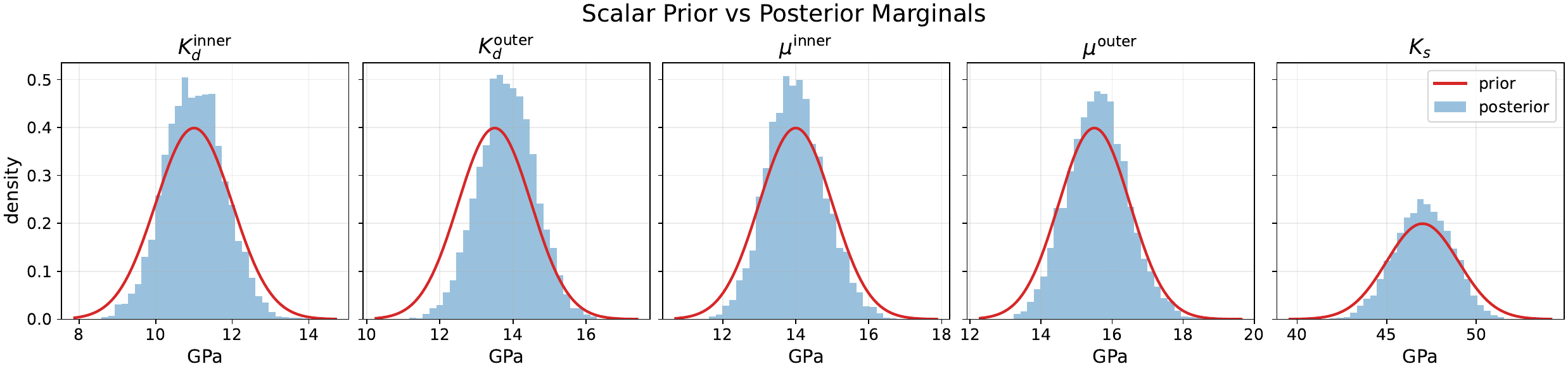}
\caption{Prior-to-posterior comparison for the scalar mechanical parameters}
\label{fig:tsx2-scalar-marginals}
\end{figure}

The scalar parameters are summarized in Figure~\ref{fig:tsx2-scalar-marginals},
where the posterior marginals are compared directly with the Gaussian
priors. The posterior means are $K^{\mathrm{inner}}_{d}=11.02$~GPa,
$K^{\mathrm{outer}}_{d}=13.83$~GPa, $\mu^{\mathrm{inner}}=14.00$~GPa,
$\mu^{\mathrm{outer}}=15.60$~GPa, and $K_{s}=47.03$~GPa. The corresponding
posterior standard deviations are approximately $0.78$~GPa, $0.78$~GPa,
$0.80$~GPa, $0.84$~GPa, and $1.58$~GPa. These values remain
close to the prior means, but the posterior marginals are visibly
sharpened relative to the prior densities, indicating moderate information
gain in the scalar mechanical parameters.

\paragraph{~}

\paragraph{~}

\paragraph{Spatial posterior fields in the logarithmic parameterization.}

\begin{figure}[H]
\includegraphics[width=0.9\textwidth]{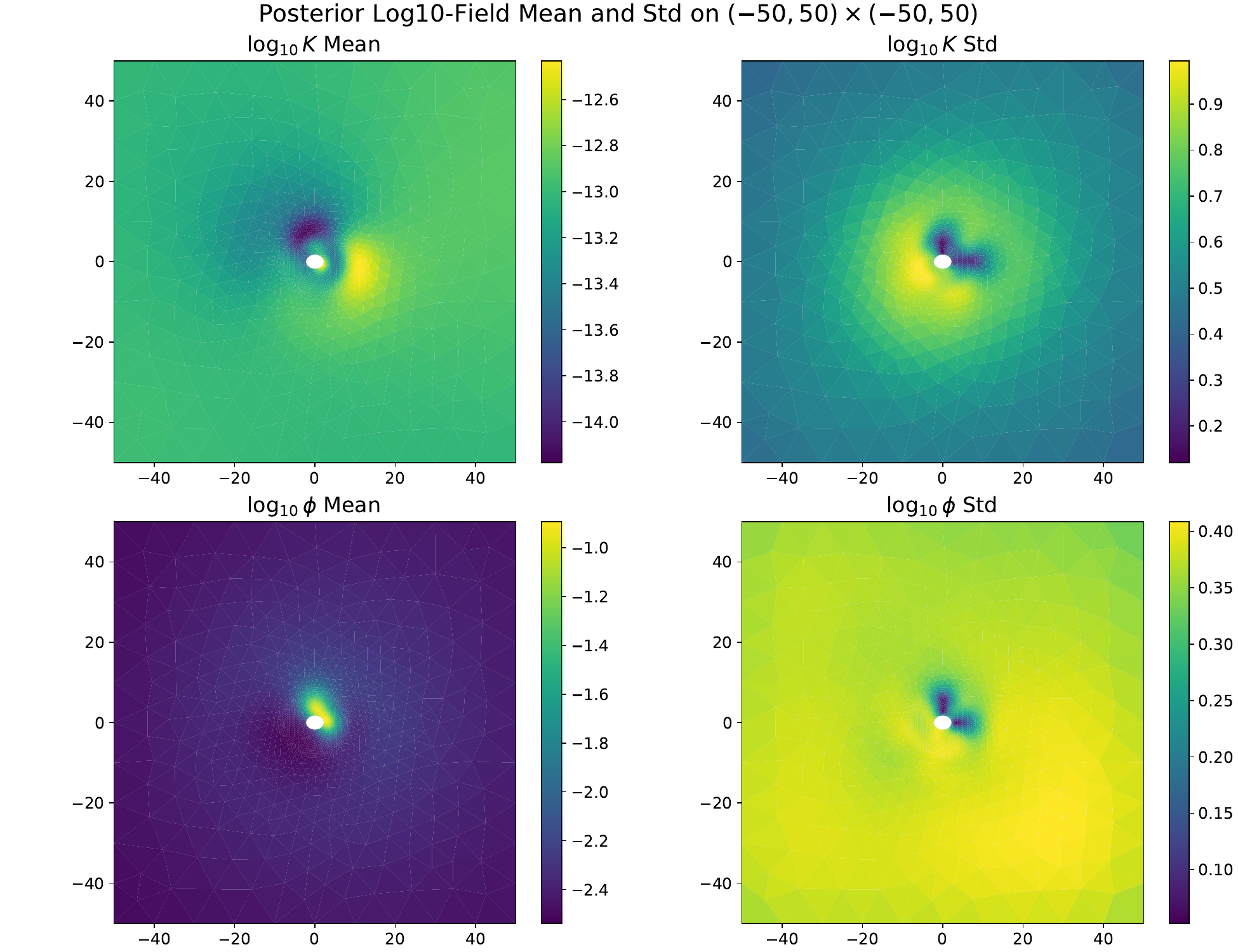}
\caption{Posterior mean and posterior standard deviation of $\log_{10}K$ and
$\log_{10}\phi$ on the full computational domain}
\label{fig:tsx2-log10-full}
\end{figure}

\begin{figure}[H]
\includegraphics[width=0.9\textwidth]{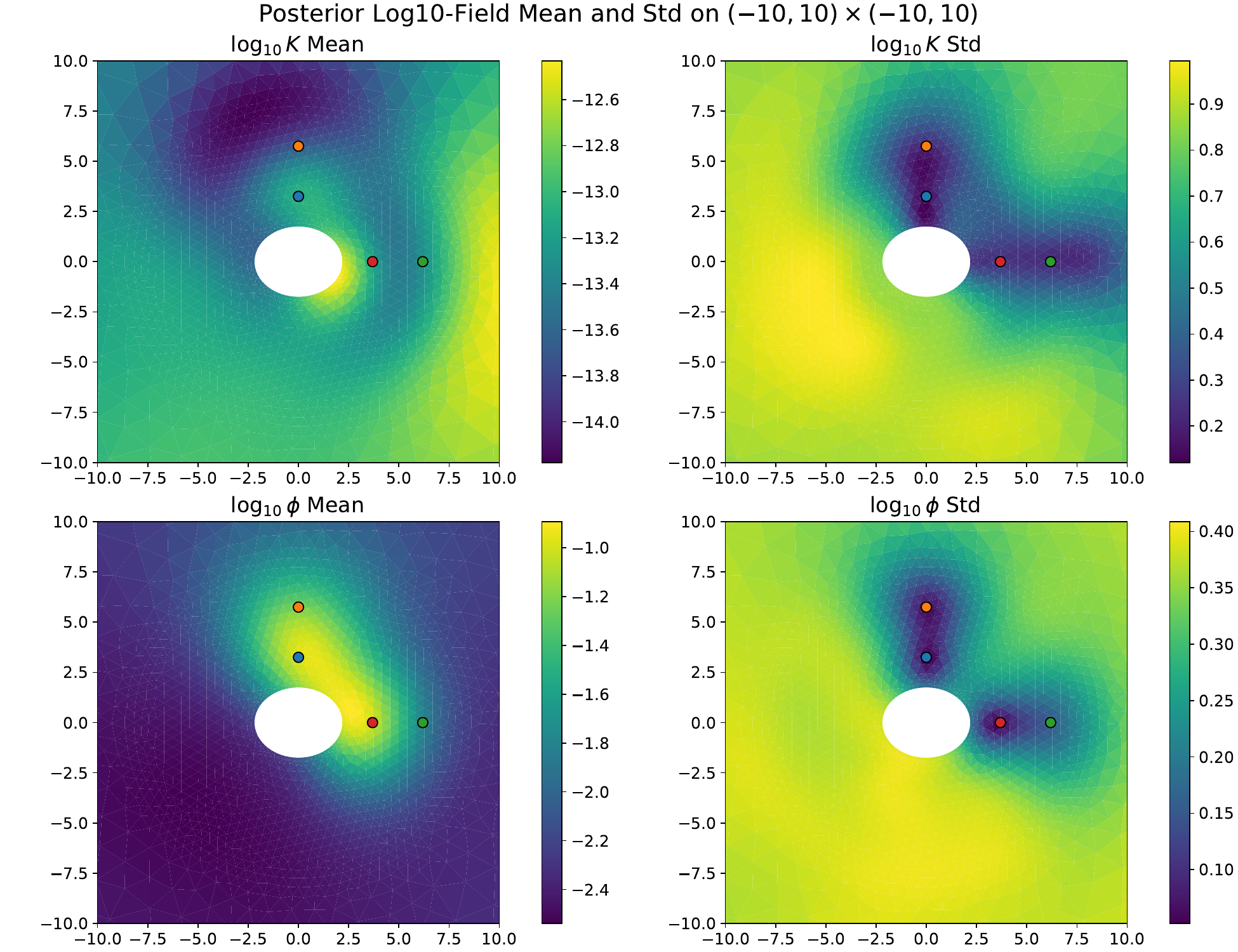}
\caption{Posterior mean and posterior standard deviation of $\log_{10}K$ and
$\log_{10}\phi$ in a tunnel-centered cutout}
\label{fig:tsx2-log10-cutout}
\end{figure}

Because the parameterization is built in terms of KL expansions of
the logarithmic fields, the most direct spatial interpretation is
obtained from the posterior moments of $\log_{10}K$ and $\log_{10}\phi$.
Figures~\ref{fig:tsx2-log10-full} and \ref{fig:tsx2-log10-cutout}
show the posterior mean and posterior standard deviation of these
fields on the full computational domain and in a local tunnel-centered
cutout. The figures show significant spatial heterogeneity in both
the posterior mean and the posterior uncertainty. The cutout view
makes the near-tunnel spatial structure easier to inspect and shows
that the posterior information depends strongly on the location of
the control points.

\paragraph{Best fit.}

\begin{figure}[H]
\centering \includegraphics[width=0.92\textwidth]{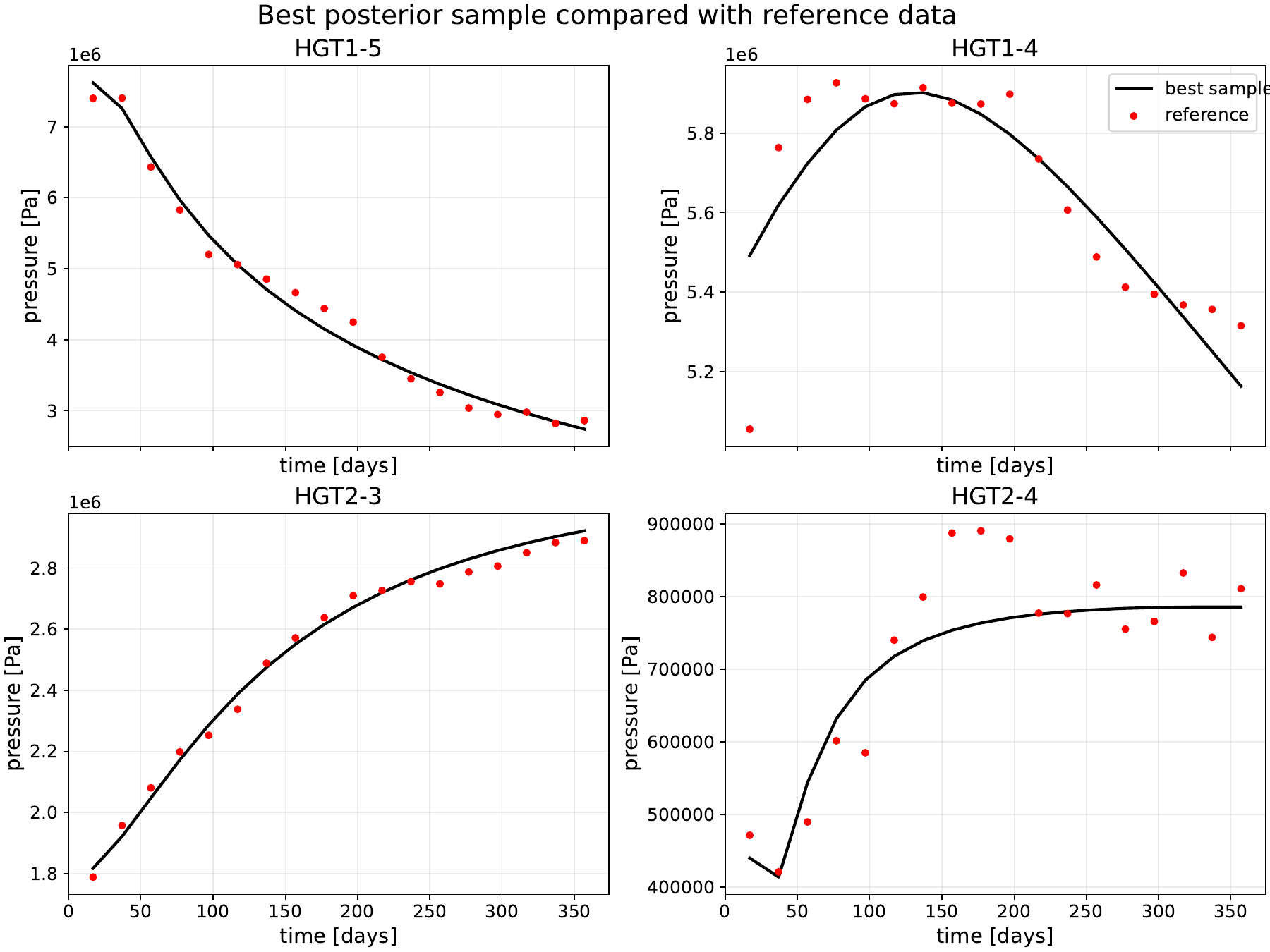}
\caption{Best posterior sample compared with the reference data}
\label{fig:tsx2-best-fit}
\end{figure}

Finally, Figure~\ref{fig:tsx2-best-fit} shows the response of the
sample with the largest data likelihood among the collected samples
against the observed data~$y\in\mathbb{R}^{72}$, i.e. the pore-pressure
time series. Searching for a best fitting sample is not the primary
outcome of the Bayesian inverse analysis. Nevertheless, it provides
a useful diagnostic: it demonstrates that the posterior support contains
parameter configurations capable of reproducing the observation trends
reasonably well.

\section{Conclusions}

This paper studied delayed-acceptance sampling with Hamiltonian proposal
subchains for Bayesian inverse problems with expensive forward models.
The method combines surrogate-based screening, Hamiltonian proposals,
and finite subchains within an exact delayed-acceptance correction,
while the surrogate is updated only during burn-in and then fixed
for production sampling.

The practical advantage comes from using cheap surrogate derivatives
obtained through automatic differentiation of the MLP surrogate to
build longer and more informative first-stage moves. At the same time,
neural-network derivatives may be less stable, especially during training,
which is one reason why the subchain construction is useful as a robust
practical variant.

The TSX benchmark allowed us to test the sampling framework in a realistic
PDE-based setting. The resulting DAHMC sampler was substantially more
efficient than the basic MH baseline (more than 60 times) and also
clearly improved over the pCN-subchain variant (about 3.8 times).
However, the TSX forward model is not expensive enough to fully expose
the potential of surrogate-based acceleration, so the gains observed
here should be interpreted as conservative. Future work should therefore
include more expensive forward models, stronger diagnostics, and a
deeper study of adaptive surrogate model updates.

\clearpage

\section*{Acknowledgements}

\begingroup\hypersetup{urlcolor=black}\urlstyle{same}This work was supported by the European Union under Grant Agreement no 101166718 (EURAD-2 project).\\
The research was co-funded by the European Union under the project INODIN, no. \url{CZ.02.01.01/00/23_020/0008487}.\\
This article was co-funded by the European Union under the REFRESH - Research Excellence For REgion Sustainability and High-tech Industries project number \url{CZ.10.03.01/00/22_003/0000048} via the Operational Programme Just Transition.\endgroup

\bibliographystyle{plain}
\phantomsection\addcontentsline{toc}{section}{\refname}\bibliography{thesis,own,literature,sota}

\end{document}